\documentstyle[12pt]{article}
\font\sevenrm=cmr7

\def\be{\begin{equation}}
\def\ee{\end{equation}}

\def\gsim{\mathrel{%
\rlap{\raise 0.511ex \hbox{$>$}}{\lower 0.511ex
\hbox{$\sim$}}}}

\def\lsim{\mathrel{
\rlap{\raise 0.511ex \hbox{$<$}}{\lower 0.511ex
\hbox{$\sim$}}}}

\def\up#1{\raise 1ex\hbox{\sevenrm#1}}

\def\build#1_#2^#3{\mathrel{
\mathop{\kern 0pt#1}\limits_{#2}^{#3}}}

\def\Fc{{\cal F}}
\def\Hc{{\cal H}}

\def\Nc{{\cal N}}
\def\Rc{{\cal R}}

\font\tenbb=msym10
\font\sevenbb=msym7
\font\fivebb=msym5
\newfam\bbfam
\textfont\bbfam=\tenbb \scriptfont\bbfam=\sevenbb
\scriptscriptfont\bbfam=\fivebb
\def\bb{\fam\bbfam}

\def\Rb{{\bb R}}

\def\D{\Delta}
\def\G{\Gamma}
\def\Om{\Omega}
\def\Si{\Sigma}

\def\a{\alpha}
\def\b{\beta}
\def\d{\delta}
\def\g{\gamma}
\def\k{\kappa}
\def\lb{\lambda}
\def\om{\omega}
\def\s{\sigma}
\def\t{\theta}
\def\ve{\varepsilon}
\def\vp{\varphi}
\def\z{\zeta}

\def\apx{\approx}
\def\eqv{\equiv}
\def\ify{\infty}
\def\lgl{\langle}
\def\nb{\nabla}
\def\ne{\not=}
\def\part{\partial}
\def\pto{\propto}
\def\rgl{\rangle}
\def\sbs{\subset}
\def\sm{\simeq}
\def\ts{\times}

\def\Lra{\Leftrightarrow}
\def\ra{\rightarrow}

\begin{document}

\hfill {\small IHES/P/96/41}

\hfill {\small LPTENS 96/42}

\vglue 3cm

\centerline{\LARGE \bf ASPHERICAL ~ GRAVITATIONAL}

\vskip .3truecm
\bigskip

\centerline{\LARGE \bf MONOPOLES}

\vglue 2.5cm

\centerline{{\large Alain~ CONNES}$^{\ a}$,~~ {\large Thibault~ 
DAMOUR}$^{\ a,b,}$\footnote{E-mail : damour@ihes.fr},~~
{\large Pierre~ FAYET}$^{\ c,}$\footnote{E-mail :
fayet@physique.ens.fr}}

\vglue 2.5cm

\begin{itemize}
\item[$^{a}$] {\footnotesize Institut des Hautes Etudes
Scientifiques, 91440 Bures-sur-Yvette, France}
\end{itemize}

\begin{itemize}
\item[$^{b}$] {\footnotesize DARC, Observatoire de Paris-CNRS,
92195 Meudon, France}
\end{itemize}

\begin{itemize}
\item[$^{c}$] {\footnotesize Laboratoire de Physique
Th\'eorique de l'Ecole Normale Sup\'erieure\footnote{Unit\'e
Propre du Centre National de la Recherche Scientifique,
associ\'ee \`a l'Ecole \linebreak Normale Sup\'erieure et \`a
l'Universit\'e de Paris-Sud}, \\
24 rue Lhomond, 75231 Paris Cedex
05, France}
\end{itemize}

\newpage

\begin{abstract}

\vskip .3truecm

We show how to construct {\it non-spherically-symmetric}
extended bodies of uniform density behaving exactly as
pointlike masses. These ``gravitational monopoles'' have the
following equivalent properties: (i) they generate, outside
them, a spherically-symmetric gravitational potential $\,M/\vert \,
{\bf x} - {\bf x}_{\rm O} \vert\,$; (ii) their interaction
energy with an external gravitational potential $\,U({\bf
x})\,$ is $\,-\,M\,U({\bf x}_{\rm O})\,$; and (iii) all their multipole
moments (of order $\ell \geq 1$) with respect to their center of mass
${\rm O}$ vanish identically. The method applies for
any number of space dimensions. The free parameters entering the
construction are: (1) an arbitrary
surface $\Si$ bounding a connected open subset $\,\Om\,$ of
$\Rb^3$; (2) the arbitrary choice of the center of mass ${\rm O}$
within $\,\Om\,$; and (3) the total volume of the body. 
An extension of the method allows one to construct 
homogeneous bodies which are {\it gravitationally equivalent~} 
(in the sense of having exactly the same
multipole moments) to any given body.

\smallskip

Though our method generally assumes that the
domain $\,\Om\,$ is boun\-ded (which leads to
bounded monopoles with {\it closed internal cavities}), it can also
generate unbounded monopoles, with exponentially-decreasing
thicknesses at infinity, and cylindriclike internal cavities
reachable from infinity. This may be useful
for optimizing the shape of test masses in high-precision
Equivalence Principle experiments, such as the planned
Satellite Test of the Equivalence Principle (STEP). By
suppressing the couplings to gravity gradients, one can
design, with great flexibility in the choice of shapes,
differential accelerometers made of nested bodies, which are
(exactly or exponentially) insensitive to all external gravitational
disturbances. Alternatively, one can construct nested bodies 
of arbitrary densities
having identical (or proportional) sequences of multipole moments, thereby also
suppressing
any differential acceleration caused by external gravity gradients.

\end{abstract}

\section{INTRODUCTION.}

The Equivalence Principle states that test bodies of different
compositions must ``fall'' exactly in the same way in an
external gravitational field. This fundamental principle,
which lies at the basis of Einstein's theory of gravitation,
deserves to be tested with the utmost precision available.
Testing the Equivalence Principle is also the most sensitive
way to search for hypothetical non-electromagnetic long range
forces. Such forces could be due to the exchanges of new spin-1
or spin-0 particles (as suggested in some supersymmetric or
superstring theories \cite{F86,DP94}), whose existence
might be required for a consistent theory of quantum gravity.
Because of their composition-dependence\footnote{New
interactions due to spin-1 exchanges are generally expected to
act on a linear combination of baryon and lepton numbers, or
more specifically on their difference $\,B-L\,$, in grand-unified
theories \cite{F86}. The spin-0 exchange forces suggested by
string theory are expected to act effectively on a linear
combination of $\,B\,$, $\,L\,$ and of the nuclear electrostatic
energy \cite{DP94}. New spin-1 or spin-0 induced forces could
also act, in addition, on particle spins.}, these forces
would, by superposing their effects to those of gravitation,
lead to (apparent) violations of the Equivalence Principle.

\medskip

Improving upon the results of the classic E\"otv\"os and Dicke
experiments, recent laboratory experiments have checked the
validity of the Equivalence Principle at the $\,\apx \, 3\ts
10^{-12}\,$ level \cite{A90}. In addition, the Lunar Laser
Ranging experiment has established that the Earth and the Moon
``fall'' towards the Sun with equal accelerations, to within
$\,\apx \, 5\ts 10^{-13\,}$ \cite{LLR}. A proposed satellite
experiment known as STEP (Satellite Test of the Equivalence
Principle) \cite{W74}, presently studied by ESA, CNES and
NASA \cite{STEP93,STEP96,CNES,NASA}, aims at improving 
the sensitivity down to the $\,\apx \, 10^{-17}\,$ level.
\medskip

The STEP experiment consists essentially of measuring the
differential accelerations between several pairs of nested
cylindrically-symmetric coaxial test masses of different
compositions freely falling within a ``drag-free'' satellite
in low orbit around the Earth. At the impressive sensitivity 
level of $\,10^{-17}$, one has to worry about many possible
disturbances. In particular, the coupling to external gravity
gradients of the two differently-shaped test masses making up
one differential accelerometer causes a parasitic differential
acceleration even if the Equivalence Principle holds exactly.
The origin of this disturbance is that an extended body is
generally not equivalent to a pointlike mass in its
interaction with an external gravitational field. Instead, the
gravitational structure of an extended body can be described
by the infinite set of its multipole moments, starting with
the monopole moment -- identical to the total mass. Choosing as origin
the center of mass of the body ensures the vanishing of its
dipole moment $(\ell =1)$ \footnote{The common vanishing of
the dipole moments of two nested masses -- i.e. the fact that their 
two centers of mass can be made to coincide -- is 
realized to a high precision in the STEP experiment by
measuring the effect of the Earth's gravity gradient and
positioning the masses so as to eliminate this effect.}, and
allows one to describe its extension-dependent interactions
with an external gravitational field in terms of its successive multipole
moments $\,Q_{\ell}\,$ of order $\,\ell \geq 2\,$. The existence of a
moment of order $\,\ell\,$ leads, in the presence of a disturbing mass
at distance $\Rc$, to an extension-dependent acceleration
$\,\pto \, Q_{\ell} / \Rc^{\,\ell +2}\,$.

\medskip

These parasitic accelerations are a serious concern in the STEP
experiment. The way this problem is currently addressed is
multi-pronged. First, while the inner mass $M^{\,\rm in}$ is
shaped as a straight cylinder with vanishing quadrupole moment
$Q_2^{\,\rm in} \!=\! 0$, but non-zero (even) higher moments
$Q_4^{\,\rm in}$, $Q_6^{\,\rm in} , \ \ldots\,$, the outer mass $M^{\rm \,
out}$ is shaped as a ``belted cylinder'' \cite{Speake,STEP93,L93} having a
vanishing quadrupole moment, $Q_2^{\,\rm out} \!= \!0$, and a non-zero hexadecapole
moment chosen so that $\,Q_4^{\,\rm out} / M^{\,\rm out} = Q_4^{\,\rm in} /
M^{\,\rm in}\,$. The latter matching condition for the reduced hexadecapole moments
$\,Q_4 / M \,$ ensures the cancelling of the
differential acceleration $\,\pto 1/\Rc^6\,$ induced by the two
$\,Q_4$'s. The remaining free parameters are optimized so as to
minimize the effects of the non-matched higher multipole
moments \cite{L93,TW96}. Second, from the calculation
of the residual sensitivity of differential accelerometers to
external masses, one deduces constraints on the allowed motion
of nearby masses. In particular, one has still severe
constraints on the amplitude of any ``helium tide'' in the
spacecraft's dewar surrounding the test masses. One must then
design the dewar so as to meet these constraints.

\medskip

In this paper, we propose a radical solution to the
problem of the sensitivity to external gravity gradients. We
show how to construct extended homogeneous bodies of very
general shapes\footnote{This very large flexibility in
the choice of the shapes of our monopoles could make
 them of practical utility in Equivalence
Principle tests. The obvious solution of 
spherically-symmetric test masses was considered by the STEP team
\cite{STEP93} but rejected because of the unsurmountable
practical difficulties of building nested spherical test masses with
independent suspension and sensing systems.} behaving {\it
exactly as pointlike masses} in any external gravitational
field. The existence of such {\it aspherical gravitational
monopoles} has been made plausible by Barrett \cite{B89} who
found approximate numerical solutions to this problem. Here 
we shall prove the existence of such
objects by providing an explicit method for constructing them\footnote{
After submitting this work for publication we learned of an unpublished 
work by Barrett \cite{barrett89}, in which he proposed the same construction 
method (for monopoles with boundaries having the topology of a sphere) without, 
however, developing it and studying its consequences.}.
Our construction allows us to have a clear control of the
large flexibility in the shapes of such
homogeneous\footnote{By contrast, the existence of
{\it inhomogeneous} aspherical monopoles is not surprising as
one can smoothly deform, within some sphere, the potential
generated by a homogeneous ball, and then {\it define} the density
``generating'' this deformed potential (which remains proportional to 
$\,1/r\,$ outside this sphere) by Poisson's equation.} monopoles.

\medskip

An extension of our method allows one to construct 
{\it gravitationally similar} bodies (of equal or different densities),
i.e. bodies having the same infinite sequences of reduced multipole 
moments $\,Q_{\ell}/M\,$.
Two such bodies experience exactly the same acceleration in the presence of 
arbitrary external gravity gradients, as soon as their centers of mass are 
made to coincide.

\medskip

As far as we know, the specific problem of the construction of
aspherical monopoles has not been dealt with in the large
literature on potential theory. (For an entry into this
literature see e.g. Ref.~\cite{Brodsky}.) In fact, there are
theorems in inverse potential theory proving, {\it under
certain assumptions}, the {\it inexistence} of such monopoles.
In particular, a famous theorem of Novikov \cite{N38} states
that, if we know the value $\,\rho_{\circ}\,$ of the supposedly 
uniform density, the exterior
potential generated by a {\it starlike}\footnote{An open set
$\,\Om\,$ in $\Rb^N$ is said to be starlike with respect to a
center ${\rm O}$ if $\,\Om\,$ contains the straight segment joining
${\rm O}$ to any point of $\,\Om\,$.} 
open set $\,\Om\,$ (with known
center ${\rm O}$) determines uniquely that domain. A corollary
of this theorem is that a homogeneous starlike monopole must be
a ball. We shall in fact prove below that this inexistence of
aspherical homogeneous monopoles holds under the much weaker
assumption that the open set $\,\Om\,$ be connected and have a
connected boundary $\,\Si = \part \, \Om \,$. 
I.e., {\it aspherical solid homogeneous monopoles cannot exist} 
(this excludes, for
instance, monopoles in the form of solid tori)\footnote{
After submitting this work for publication we learned from a referee 
of an unpublished work by Barrett \cite{barrett90}, in which he established 
this result by a different, more involved method.}. 

\medskip

This no-go theorem, however, does not exclude aspherical monopoles 
-- either simply
connected\footnote{In Ref.~\cite{N38}  one finds,
without proof (on the ground that it is ``easy to prove''!),
the {\it incorrect}~ proposition that a homogeneous {\it simply
connected}~ monopole (with $C^2$ boundary) must be a ball.}
or not --  having a {\it disconnected}~ boundary $\,\Si = $ \linebreak 
$\Si_{\,\rm ext} \,
  \cup \ \Si_{\,\rm int}\,$. 
Indeed, we shall construct such homogeneous aspherical
mono\-poles, which have a
closed (in the sense of being unreachable from infinity) internal cavity.

\medskip

Our construction allows us to choose arbitrarily, either the
outer boundary, or the inner boundary (surface of the internal
cavity) of the monopole. The topology of these boundaries does
not need to be that of the sphere $S^2$. For instance, the outer boundary
can be toroidal. In the simplest\footnote{As discussed below, our basic method 
allows generalizations and combinations which lead to monopoles of very  
general shapes, having, for instance, as many holes as desired.}
bounded monopoles we construct, the center of mass 
-- which is the center of symmetry of the generated external gravitational 
field -- can be systematically
located within the internal cavity. 
This allows for a second monopole to be placed inside the first one,
with the same center of mass. 
But this inner monopole is then unreachable
from the outside by a continuous path. It means that, in
practical applications such as STEP, where one needs to
control and sense the position of nested monopoles, one can
only use {\it approximate} monopoles.

\medskip
Interestingly, we can get practically useful shapes
by considering a limit where the boundaries are not compact, but
extend to infinity. In particular, we shall discuss in detail
the case (directly relevant for STEP) where both the 
internal and external
boundaries asymptotically approach a straight cylinder at
infinity. The thickness of the corresponding 
monopole, around this limiting straight cylinder, then tends {\it
exponentially} to zero at infinity. By cutting
off such a monopole, one can construct a finite, cylinderlike
object whose  couplings with gravity gradients are suppressed with
an exponentially good precision.

\medskip

After formulating the problem in technical terms and proving
the inexistence of ``solid'' aspherical homogeneous monopoles in Section
2, we present our method for constructing ``hollow'' aspherical monopoles
in Section 3. We also indicate how, by a generalization of this method, 
one can construct homogeneous bodies which are ``gravitationally equivalent'', 
or 
``gravitationally similar'' to any given body, or collection of bodies.
In Section 4, we build our intuition for the
resulting shapes of such monopoles by studying, to lowest
order, some simple examples, in particular the thin cylinder. 
In Section 5, we discuss some of
the mathematical properties of the construction, present a
detailed analysis of the perturbations of a monopole, 
and outline the program of a
rigourous mathematical proof of the existence of such monopoles.
Physical considerations on the growth of monopoles, 
including a discussion of different
practical methods for realizing them, are presented in
Section 6.

\section{Definitions of monopoles 
and inexistence of ``solid'' aspherical homogeneous monopoles.}

\subsection{Equivalent definitions of monopoles.}

Let $\,\rho \, ({\bf x})\,$ denote a volumic mass density in
three-dimensional space. For simplicity, we start by assuming
that $\,\rho \,({\bf x})\,$ has a compact support. The gravitational
potential generated by $\,\rho \,({\bf x})\,$ reads (with our sign convention,
and setting Newton's constant $\,G_N\,$ to unity)
\be
U_{\rho}\, ({\bf x})\ \, = \,\ \int d^3 \, {\bf x}' \ \, \frac{\rho\, ({\bf
x}')}{\vert {\bf x} - {\bf x}' \vert} \ \ . \label{eq:2.1}
\ee
The interaction energy of this mass distribution $\,\rho\,({\bf x})\,$ in an
external gravitational potential $\,U_{\rm ext} ({\bf x})\,$ is
\be
E_{\rho} \ \,=\ \, - \,\int d^3 \, {\bf x} \ \, \rho\, ({\bf x})\  \, U_{\rm
ext} ({\bf x}) \ \  . \label{eq:2.2}
\ee
Both $U_{\rho}$ and $E_{\rho}$ can be skeletonized by
introducing the multipole moments of the distribution $\rho$.
They can be described by symmetric trace-free (STF) tensors
of order $\,\ell \geq 0\,$,
\be
Q_{\rho ,\,{\bf x}_{\rm O}}^{\,i_1 i_2 \ldots i_{\ell}} \ =\ {\rm
STF}_{i_1 \ldots i_{\ell}} \int d^3 \, {\bf x} \ \ \rho ({\bf
x}) \ \, (x^{i_1} - x_{\rm O}^{i_1}) \,  (x^{i_2} - x_{\rm O}^{i_2})
\,\ldots\, (x^{i_{\ell}} - x_{\rm O}^{i_{\ell}}) \ \  ,
\label{eq:2.3} 
\ee
where $\,{\rm STF}_{i_1 \ldots i_{\ell}}$ denotes a symmetric
trace-free projection over the indices $i_1 \ldots
i_{\ell}\,$\footnote{E.g., ${\ \rm STF}_{ij}\  X^i X^j = X^i X^j
-\frac{1}{N}\, \d^{ij} \, X^s X^s\,$, 
\ \ \ \ \ \ \ \ \ \ \ \ \ \ \ \ \ \ \ \ \ \break 
${\rm STF}_{ijk} \ X^i X^j X^k \ =\ 
X^i X^j X^k - \frac{1}{N+2}\ X^s X^s \ (\d^{ij} X^k + \d^{jk}
X^i + \d^{ki} X^j)\,$, etc., in $N$ space dimensions. \vskip .2truecm
},
and ${\bf x}_{\rm O}$  the coordinates of some
origin ${\rm O}$ in space. An equivalent description uses the
spherical harmonics and reads, in the three dimensional space: 
\be
Q_{\rho,\,{\bf x}_{\rm O}}^{\ell m} \ =\  \int d^3 \, {\bf x}\  \  \rho \,
({\bf x}) \ \ r^{\ell} \ \,Y_{\ell m} (\,\t ,\,\vp) \ \  ,
\label{eq:2.4} 
\ee
where $\,x^i - x_{\rm O}^i = (\,r \,\sin \t \,\cos \vp , \ r \,\sin \t\, \sin \vp
,\ r\, \cos \t \,)$.

\medskip

The zeroth order multipole moment (or monopole moment) is simply 
the total mass of the distribution, $\,M = \int d^3 \,
{\bf x} \ \, \rho \, ({\bf x})\,$. We shall always fix the origin
${\rm O}$ at the center of mass of the distribution
$\rho$, i.e. at the point such that the dipole moment 
$\,Q_{\rho ,{\bf x}_{\rm O}}^{\,i} = 
\int d^3 \, {\bf x} \,\ \rho \, ({\bf x})\ \, (x^i - x_{\rm O}^i)\,$
 vanishes identically.

\bigskip

A (bounded) gravitational monopole can then be equivalently defined as a
distribution $\,\rho\,$ satisfying either: 

({\bf i}) the external
potential generated by $\,\rho\,$ is spherically symmetric:
\be
U_{\rho} ({\bf x})\  = \ \frac{M}{\vert {\bf x} - {\bf x}_{\rm
O} \vert} \ \ \ \ \ \hbox{for} \ {\bf x} \ \hbox{in the
exterior\footnote {}
of the support of} \ \,\rho \, ; \label{eq:2.5}
\ee
or \footnotetext{If the support of the bounded
distribution $\,\rho\,$ divides space in several regions, we mean
by ``exterior'' the outside region, connected to infinity.}
({\bf ii}) the
interaction energy of $\,\rho \, ({\bf x})\, $ in any external gravitational 
potential 
$U_{\rm ext}\, ({\bf x})\,$ is the same as for a 
point mass located at the center of mass $\,{\bf x}_{\rm O}$:
\be
E_{\rho} \ =\  -\ M \ \, U_{\rm ext} ({\bf x}_{\rm O}) \ \,  ;
\label{eq:2.6} 
\ee
or ({\bf iii}) all multipole moments (of order $\ell \geq 1$)
vanish identically:
\be
Q_{\rho,\,{\bf x}_{\rm O}}^{\,i_1 i_2 \ldots i_{\ell}} \,=\, 0 \quad \ \ 
\hbox{for} \quad \ell \geq 1 \  . \label{eq:2.7}
\ee

\medskip

The equivalence between (i) and (ii) follows from considering
the external distribution $\,\rho_{\rm ext}\,$ generating 
the gravitational potential $\,U_{\rm
ext}\,$. The corresponding interaction energy is 
$\,E_{\rho} [U_{\rm ext}] = -\int d^3 \, {\bf x}
\, \rho  ({\bf x}') \,  \rho_{\rm ext} ({\bf x}) \, \vert {\bf
x} - {\bf x}' \vert^{-1}$ $ 
= \,-\, \int d^3 \, {\bf x} \ \, U_{\rho}
({\bf x}) \ \, \rho_{\rm ext} ({\bf x})\ $,
to be compared with
$\ -\, M \  U_{\rm ext} ({\bf x}_{\rm O})  \,
= \linebreak -\, \int d^3 \, {\bf x} \ \, 
\frac{M}{\vert {\bf x} - {\bf x}_{\rm O} \vert}
  \ \, \rho_{\rm ext} ({\bf x})\  $. Demanding the equality (\ref{eq:2.6})
of these two expressions for any $\,\rho_{\rm ext} ({\bf x})\,$ 
is equivalent to Equation (\ref{eq:2.5}).

\medskip

The equivalence of either (i) or (ii)
with (iii) follows from expanding
$\vert {\bf x} - {\bf x}' \vert^{-1}$ in
Equation (\ref{eq:2.1}) in powers of $\,{\bf x}'-{\bf x}_{\rm O}\,$, which 
ultimately yields,
for ${\bf x}$ in the exterior of $\,{\rm supp}\ \rho\,$:
\be
U_{\rho} ({\bf x}) \ \, =\ \, \sum_{\ell =0}^{\ify} \
\frac{(-)^{\ell}}{\ell \,!} \ \ Q_{\rho ,{\bf x}_{\rm O}}^{\,i_1
\ldots i_{\ell}} \ \ \ \part_{i_1 \ldots i_{\ell}} \ 
\frac{1}{\vert {\bf x} - {\bf x}_{\rm O} \vert} \ \  , 
\label{eq:2.8}
\ee
and from expanding $\,U_{\rm ext}({\bf x})\,$ in Equation
(\ref{eq:2.2}) in powers of  $\,{\bf x} - {\bf x}_{\rm O}\,$, which yields:
\be
E_{\rho} \ \, = \ \, -\ \sum_{\ell =0}^{\ify} \ \frac{1}{\ell \, !} 
\ \ Q_{\rho
,{\bf x}_{\rm O}}^{\,i_1 \ldots i_{\ell}} \ \ \ \part_{i_1 \ldots
i_{\ell}} \ U_{\rm ext} ({\bf x}_{\rm O}) \ \  . \label{eq:2.9}
\ee

\medskip

The properties ({\bf i}) -- ({\bf iii}) are also equivalent to:

({\bf iv}): the spatial Fourier transform $\,\hat \rho\,({\bf k})= \int d^3
{\bf x} \ e^{\,-\,i\,{\bf k} \cdot {\bf x}} \ \rho \,({\bf x})\,$  
(choosing here ${\rm
O}$ as the origin) may be expanded (around  $\,{\bf k} = {\bf 0}$) as 
\be
\hat \rho\,({\bf k}) \ = \ \hat \rho\,({\bf 0}) \ +\  {\bf k}^2 \ f({\bf k}) \ ,
\ee
as we shall discuss later (cf. subsection 4.2).

\bigskip

Essentially all our results straightforwardly extend to any number of
space dimensions $\,\,N\geq 2\,$. It suffices to replace the basic
Newtonian potential $\,r^{-1}$ by $\ (N-2)^{-1} \ r^{-(N-2)}\ $
when $\,N\ne 2$ (and by $\,\ln 1/r\,$ for $\,N=2$), and the coefficient
$\,4\,\pi\,$ in the Poisson equation
\be
\D \ U_{\rho} ({\bf x})\  = \ -\ 4\,\pi \ \rho\, ({\bf x})
\label{eq:2.10} 
\ee
by the surface of the unit sphere in $\Rb^N$, $\,\om_N =
2\ \pi^{N/2} /\, \G (N/2)\,$. To simplify the discussion we shall
phrase our results by assuming $\,N=3$. 

\medskip

An {\it homogeneous} monopole is defined by a density $\,\rho\,$
uniform within some domain\footnote{By ``domain'' we
mean a {\it connected} open subset of $\Rb^N$.} $\,\Om\,$ of
$\Rb^3$. For simplicity, we assume that $\,\Om\,$ is bounded and
admits a sufficiently regular boundary $\,\Si = \part \, \Om\,$ 
(assumptions that we shall relax later). 

\medskip

It is useful to introduce also the concepts of {\it gravitationally equivalent~} 
and {\it gravitationally similar~} distributions. Two density distributions
$\,\rho_1\,({\bf x})\,$ and $\,\rho_2\,({\bf x})\,$ will be said 
``gravitationally equivalent'' if they generate the same external gravitational 
potential, i.e. if all their multipole moments coincide:
\be
Q_{\rho_1,\,{\bf x}_{\rm O}}^{\,i_1 i_2 \ldots i_{\ell}} \ =\  
Q_{\rho_2,\,{\bf x}_{\rm O}}^{\,i_1 i_2 \ldots i_{\ell}}  \quad \ \ 
\hbox{for} \quad \ell \geq 0 \  \ .
\ee
(the choice of origin ${\rm O}$ being arbitrary).
The distributions will be said ``gravitationally similar'' 
if the external gravity potentials they generate are proportional.
This is equivalent to requiring that their ``reduced'' multipole moments coincide:
\be
Q_{\rho_1,\,{\bf x}_{\rm O}}^{\,i_1 i_2 \ldots i_{\ell}} \ / \ M_1\ \,=\,\ 
Q_{\rho_2,\,{\bf x}_{\rm O}}^{\,i_1 i_2 \ldots i_{\ell}} \ / \  M_2 \  \ .
\ee
This latter concept is invariant under a (constant) rescaling of the densities 
$\,\rho_1\,$ and $\,\rho_2\,$. In this nomenclature, a monopole can be defined 
as a distribution which is gravitationally equivalent, or similar, to a point mass.

\subsection{\sloppy Inexistence of aspherical solid homogeneous mono\-poles.}

Let us prove that if $\,\Om\,$ is a (connected) homogeneous monopole, and if
the boundary $\,\Si = \part \, \Om\,$ of the domain
$\,\Om\,$ is {\it connected}, then $\,\Om\,$ must be a solid sphere.
To do so we study the action, on the potential $\,U_{\rho} ({\bf x})\,$
generated by the monopole $\,\Om$,
of the generators of infinitesimal rotations\footnote{If 
the number of space dimensions is not $\,N = 3$, we use
\vskip -.3truecm
$$
L^{j\,k}\ = \ (x^j -x_{\rm O}^j)\ \frac{\part \ \ }{\part\,x^k}\ -\
 (x^k -x_{\rm O}^k)\ \frac{\part \ \ }{\part\,x^j}\ ,   
$$
\vskip -.1truecm
\noindent
the subsequent argument remaining the same.
\vskip .1truecm
} around ${\rm O}$,
\be
\mbox{\boldmath $L$}\ = \ ({\bf x} -{\bf x}_{\rm O})\ \times \
\frac{\part \ \,}{\part\,{\bf x}}\ \ .
\label{eq:2.11n}
\ee
The rotational symmetry of the Laplacian implies that the (Lie) derivative
of $\,U_{\rho} ({\bf x})$, say $\, \mbox{\boldmath $u$} = \ 
\mbox{\boldmath $L$} \ U_{\rho}\,$,
 satisfies everywhere, in the sense of distributions, 
\be 
\D \ \mbox{\boldmath $u$} \ = \ -\ 4\,\pi \ \mbox{\boldmath $L$} \ \rho \ ,
\label{eq:2.11}  
\ee
in which the source $\,\mbox{\boldmath $L$} \, \rho\,$ 
is
a ``single layer'' surface density distribution\footnote{To be explicit, 
we have
\vskip -.3truecm
$$
\mbox{\boldmath $L$} \ \rho \ =\ \mbox{\boldmath $ \s$} 
\ \d_{\,\Si}\ =\ -\ ({\bf x} - {\bf x}_{\rm O}) 
\ \times \ {\bf n} \ \ \rho_{\circ}\ \d_{\,\Si}   \nonumber
$$
\vskip 0truecm
\noindent
We have used the fact that the gradient of the characteristic 
function of the 
domain $\Om$ is $\,-\,{\bf n} \ \d_{\,\Si}\,$,  $\,{\bf n}\,$
being the outgoing normal to the boundary $\,\Si = \part \,\Om\,$.
$\d_{\,\Si}$ denotes the standard uniform $\d$-function
surface distribution on $\Si$ (such that $\,\int d^3 \, {\bf x} \ \,
f({\bf x}) \ \d_{\,\Si} \,=\, \int_{\Si} \, d\, \Si\ \,[f({\bf x})]_{\Si}\,$,
where $d\,\Si$ is the area element on $\Si$).}.

\medskip
 
Let us assume that 
$\,\Om\,$ defines a homogeneous monopole with a {\it connected}~ boundary 
$\,\Si = \part \, \Om\,$. This implies that $\,\Si\,$ divides space into two
 regions: the domain $\,\Omega\,$ (``interior of $\,\Si$'') and the
complement of $\,\Omega\,$ (``exterior of $\,\Si$''). 
Each component $\,u^i\,= L^i
\,U_{\rho}\,$  of the infinitesimal variation of the potential appears itself as
a potential,  generated by the single layer surface density $\,L^i\,\rho =\s^i\,
\d_{\,\Si}\,$,  distributed on the  boundary $\,\Si \,$. We shall now prove
that this surface density $\,\s_i\,$ must in fact vanish identically.
Indeed the potential $\,u^i$ it generates
vanishes everywhere in the exterior of $\,\Si\,$ (since $\,U_{\rho} ({\bf x})\,$
is rotation-invariant), and is harmonic everywhere inside it.
On the surface $\,\Si\,$ itself $\,u^i\,$ must be continuous, while
its normal derivative might a priori have a discontinuity,
equal to $\,-\,4\,\pi\,$ times 
the surface density $\,\s^i\,$. The vanishing of $u^i$ in the exterior and its
continuity across $\Si$ implies that it vanishes on $\Si$. It must then vanish
everywhere inside $\,\Si\,$ (by the uniqueness of the solution  of the Dirichlet
problem). Since it also vanishes outside, its derivative cannot have 
a discontinuity on the surface $\,\Si$, which implies that the density
$\,\s^i\,$ distributed on $\,\Si\,$ has to vanish. 
The resulting equation,
\be
\mbox{\boldmath $L$} \ \rho \ =\ {\bf 0} \ ,
\ee
expresses that the monopole $\,\Om\,$ should be spherically
symmetric\footnote{Technically, the formula of the previous
footnote shows that $\,\mbox{\boldmath $L$} \, \rho = 0\,$  implies that every
normal $\,{\bf n}\,$ to $\,\Si\,$ should pass  through ${\rm O}$, i.e. that the
connected surface $\,\Si\,$  is a sphere of center ${\rm O}$, and the monopole
$\,\Om\,$ the corresponding solid sphere.}. It must be the solid sphere limited
by the sphere $\,\Si\,$ of center $\rm O$.

\medskip

We have not restricted in advance $\,\Om\,$ to
be simply connected. Our result therefore excludes, for example, 
``solid'' tori as possible homogeneous mono\-poles. 

\medskip

The crucial assumption is that the connected domain $\,\Om\,$ have 
 a connected boundary $\,\Si\,$ (i.e. that we are dealing with
a ``solid'' monopole without any internal cavity). By contrast, in
the case where the boundary of $\,\Om\,$ is made of two
disconnected surfaces, one inside the other 
(i.e. $\,\part \, \Om = \Si_{\rm ext} \cup \,\Si_{\rm int}\,$,
as for a ``thickened'' topological sphere or topological torus, 
i.e. the volumes between two
nested surfaces having the topology of $\,S^2\,$ 
or $\,T^2\,$), or more than 
two\footnote{This is the case of a ``Swiss-cheese-like''
domain $\,\Om\,$ with a disconnected boundary 
$\,\Si = \part \, \Om = \Si_{\rm ext} \cup \{\Si_{{\rm int}\,i}\}  \,$.}, 
we have no way to conclude that the monopole 
must be spherically symmetric.
Indeed we shall see that, in such cases, one can
construct infinitely many aspherical homogeneous
monopoles.

\bigskip

Our analysis can easily be further extended to aspherical monopoles 
\linebreak
$\Om= \cup \ \Om_l\,$ that
would be constituted of several {\it disconnected} parts
 (with boundaries $\,\Si_l = \part \, \Om_l\,$). 
If the exterior boundary $\,\Si_{{\,\rm ext}\ l}\,$ of one of them
 does not enclose the center ${\rm O}$ inside it, 
and is reachable from infinity, it follows from Gauss' theorem 
that the total mass included
within it must vanish. 
As a result, {\it the exterior 
boundary of a gravitational monopole must always be connected}. 
For example a collection of solid objects 
(e.g. solid tori, even intertwined) 
cannot constitute a gravitational monopole.
If a monopole is constituted
of several components all but one should be located 
within the interior of one of them, as in the case 
of Russian dolls, for example.

\section{\sloppy Constructing aspherical (hollow) mono\-poles.}

\subsection{Thin aspherical monopoles.}

The starting point of our method is a simple consequence of a
well-known property of electrostatics (see e.g. Ref.~\cite{Jackson}).
Let $\,\Om\,$ be a domain with a connected boundary $\,\Si\,$. Let us
consider $\,\Si\,$ as a grounded conducting surface, and 
introduce a unit (positive) electric charge at some given
point ${\rm O}$ (with coordinates $x_{\rm O}^i$) within $\,\Om\,$.
The charge $+\,1$ at ${\rm O}$ induces on the surrounding
grounded conducting surface $\,\Si\,$ a (negative) charge density,
say $-\,\s$, which has the well-known property of generating a
potential which exactly {\it screens} the $\,1/\vert {\bf x} - 
{\bf x}_{\rm O} \vert\,$ potential created by the charge $\,+\,1\,$, {\it
outside} $\,\Si\,$. (This is mathematically evident from the
uniqueness of the solution of the Dirichlet problem $\,\D\,U =0\,$,
 in the complement of
$\,\Om\,$.) By reversing the sign of the surface charge density,
we conclude that the positive surface density $\,+\,\s\,$ on $\,\Si\,$
generates, outside this surface, the exactly spherical potential $\,+\,1
/ \vert {\bf x} - {\bf x}_{\rm O} \vert\,$.

\medskip

In mathematical terms, the potential generated by the charge $+\,1\,$
introduced at ${\rm O}$ is the Dirichlet Green function of the domain 
$\,\Om\,$,
 $\,G({\bf x} ,{\bf x}_{\rm O})$, solution of
\be
\D_x \ G ({\bf x} ,{\bf x}_{\rm O})\  = \ -\ 4\,\pi \ \d ({\bf x} - 
{\bf x}_{\rm O}) \ , \label{eq:3.1}
\ee
which vanishes when $\ {\bf x} \in \Si \!= \part \, \Om\,$. Let us
define the following surface density on $\,\Si\,$
\be
\s_{\Si}^{\rm O} ({\bf x}) \ =\ -\ \frac{1}{4\pi} \ \part_n \, G
({\bf x} , {\bf x}_{\rm O}) \ , \label{eq:3.2}
\ee
where $\,\part_n\,$ denotes the outgoing normal derivative at $\,{\bf
x} \in \Si\,$. The density $\,\s_{\Si}^{\rm O}\,$ is everywhere
positive\footnote{$\,\s_{\Si}^{\rm O}\,$ is positive even if $\,\Om\,$ has a
non trivial topology, e.g. that of a multi-handled solid
torus. Indeed, $\,G({\bf x})\,$ being positive (and $\,\sm 1 / \vert 
{\bf x} - {\bf x}_{\rm O} \vert\,$) near $\,{\bf x}_{\rm O}\,$, zero
on the boundary of $\,\Om\,$, and harmonic within $\,\Om - \{ {\rm
O} \}\,$ cannot (by the maximum-minimum principle) reach
negative values within $\,\Om\,$. Therefore $\,G({\bf x})\,$ decreases
from positive to zero values when crossing $\,\Si\,$.} and
integrates to $\,+\,1\,$:
\be
\int_{\Si}\ d\,\Si \ \, \s_{\Si}^{\rm O}\  = \ -\ \frac{1}{4\pi} \,
\int_{\Si}\  d\,\Si \ \,\part_n \, G \ = \ -\ \frac{1}{4\pi} \,\int_{\Om}
d^3 \, {\bf x} \ \ \D_x \, G\  =\  +\ 1 \ . \label{eq:3.3}
\ee
Applying Green's identity
\be
\int_{\Om} \  d^3 \, {\bf x} \ \, (u \, \D \, v - v \, \D \, u) \ =
\  \int_{\Si}\ 
d\,\Si \ \ (u \ \part_n \, v - v \ \part_n \, u) \label{eq:3.4}
\ee
with $u$ equal to any external potential $\,U_{\rm ext}\,$
(satisfying $\,\D \, U_{\rm ext} = 0\,$ within $\Om \cup \Si\,$),
and $\,v({\bf x}) = G({\bf x} ,{\bf x}_{\rm O})\,$, yields
\be
U_{\rm ext} ({\bf x}_{\rm O}) \ =\  \int_{\Si}\ d\,\Si \ \ 
\s_{\Si}^{\rm O} ({\bf x}) \ \ U_{\rm ext} ({\bf x}) \ .
\label{eq:3.5} 
\ee

Equation (\ref{eq:3.5}) expresses precisely the property
(\ref{eq:2.6}) of a monopole: the surface layer with density
$\,\s_{\Si}^{\rm O}\,$ interacts with any external potential as if
it were a unit point mass located at ${\rm O}$. (The more
``active'' definition (\ref{eq:2.5}) is obtained by taking
$\,U_{\rm ext} ({\bf x}) = \vert {\bf x} - {\bf x}_{\rm
P}\vert^{-1}\,$ for some external point $P$.) In the mathematical
literature, the surface distribution $\,\s_{\Si}^{\rm O} \ d\,\Si\,$
is called the ``harmonic measure'' of $\,\Si\,$ with respect to the
origin ${\rm O}$. Equation (\ref{eq:3.5}) says that
the value, at a given point ${\rm O} \in \Om$, of any harmonic
function can be written as a weighted average of the values
taken by this harmonic function on the boundary of $\,\Om\,$. In other words,
$\s_{\Si}^{\rm O} ({\bf x})$, if it is known for all points ${\rm O} \in
\Omega$, gives, via Eq.~(\ref{eq:3.5}), the explicit (unique) solution of the
Dirichlet problem: determining a harmonic function in $\,\Omega\,$ from its 
values on $\,\Si = \part \,\Omega$.

\medskip

In words, the consideration of the surface density
$\,\s_{\Si}^{\rm O}\,$ solves the problem of constructing
infinitely thin aspherical monopoles. Note that $\,\Si\,$ does not
need to have the topology of the sphere. It can have the
topology of a torus, or of a more complicated surface with
many handles. As long as $\,\Si\,$ is the connected boundary of a
(sufficiently regular\footnote{We refer to the literature on
potential theory, e.g. Ref.~\cite{Doob}, for the characterization
of the domains admitting a Green function and a harmonic
measure. We shall briefly discuss later what happens when
$\,\Si\,$ is not smooth.}) connected open set $\,\Om\,$, one can
construct a thin monopole having the shape of $\,\Si\,$. It is
surprising, but true, that one can choose arbitrarily a
complicated multi-handled shape $\,\Si\,$, and an arbitrary origin
${\rm O}$ enclosed within $\,\Si\,$, and construct, by depositing
a positive surface layer $\pto \s_{\Si}^{\rm O}\,$ on $\,\Si\,$, an
object generating a potential which is exactly proportional to $\,\vert 
{\bf x} - {\bf x}_{\rm O} \vert^{-1}\,$ outside $\,\Si\,$. At this
stage, we assume that $\,\Om\,$ is bounded, but we shall later
consider the limiting case where $\,\Om\,$ becomes unbounded in a
way which ensures that the mass distribution of the monopole
falls off exponentially fast at infinity.

\subsection{Thick aspherical monopoles.}

To construct ``thick'' aspherical monopoles, it suffices to
build them by growing successive layers having a thickness
proportional to $\,\s_{\Si}^{\rm O}\,$. More precisely, let us
start with some initial closed surface $\,\Si\,$ which is the
(connected) boundary of a domain $\,\Om\,$ and let us choose, once
and for all, some origin ${\rm O}$ within $\,\Om\,$. It is
convenient to use as parameter $t$ measuring the continuous
growing of a thick monopole around $\,\Si\,$ the algebraic volume
of the constructed monopole, counted positive for monopoles
obtained by adding layers to the outside of $\,\Si\,$, and
negative when growing $\,\Si\,$ towards the inside. At each
``time'' $\,t\,$ the monopole consists of the volume $\,\Om_t\,$ 
delimited by
two boundaries: a fixed boundary $\,\Si\,$ and a moving one, say
$\,\Si_t\,$. 

\medskip

We can (locally) represent the moving boundary $\,\Si_t\,$ by equations of
the form
\be
{\bf x} \, \in \,\Si_t \ \ \,\Lra \ \ \,{\bf x} = 
{\bf X}\, (t, \hbox{\boldmath
$\xi$})\ , \label{eq:3.6}
\ee
where $\,\hbox{\boldmath $\xi$} = (\xi^1 , \xi^2)\,$ denote two
(curvilinear) coordinates on the fixed boundary $\,\Si\,$. In
mathematical terms, $\,{\bf X} (t,\cdot )\,$ is an embedding of
$\,\Si\,$ into $\Rb^3$. We define this embedding (as can always be
done) so that points with fixed coordinates {\,\boldmath $\xi\,$}
propagate orthogonally to the instantaneous $\,\Si_t\,$. The
condition of adding successive infinitesimal layers of
constant density and thickness proportional to the harmonic
measure $\,\s_{\Si_t}^{\rm O}\,$ reads
\be
\frac{\part}{\part\, t} \ {\bf X}\, (t,\hbox{\boldmath $\xi$}) \ =\ 
\s_{\Si_t}^{\rm O}\, ({\bf X} (t,\hbox{\boldmath $\xi$})) \ \ 
{\bf n}_{\Si_t}\, ({\bf X} (t,\hbox{\boldmath $\xi$})) \ ,
\label{eq:3.7}
\ee
where $\,{\bf n}_{\Si_t}\,$ denotes the outward normal to $\,\Si_t\,$.
The differential version of Equation (\ref{eq:3.7}), 
\be
\,d\,{\bf X}\ = \
\s_{\Si_t}^{\rm O} \ \, {\bf n}_{\Si_t} \ dt\,\ ,
\label{eq:3.7'}
\ee
together with
the fact that the surface integral $\ \int d\,\Si_t \ \s_{\Si_t}^{\rm O}\ $ is
$+\,1$ (Equation (\ref{eq:3.3})), makes it clear that the
infinitesimal volume between $\,\Si_t\,$ and $\,\Si_{t+dt}\,$ 
is indeed $\,dt\,$, and
that each layer $\,d\,{\bf X}\,$ adds an infinitesimal {\it
homogeneous} monopole $\,d\,\Om_t\,$ with fixed origin ${\rm O}$. (The
constant density $\,\rho_{\circ}\,$ of the monopole is set to
unity for simplicity.) 

\medskip

This can also be verified by a simple
calculation. Namely, Equations (\ref{eq:3.7}) or (\ref{eq:3.7'}) are
equivalent to requiring that the $t$-derivative of the
integral of any function $\,u({\bf x})\,$ over the volume $\,\Om_t\,$
contained between $\,\Si\,$ and $\,\Si_t\,$ be equal to the surface
integral $\,\int\, d \,\Si_t \ \s_{\Si_t}^{\rm O} \ u\,({\bf X}
(t,\hbox{\boldmath $\xi$}))\,$. If $\,u({\bf x})\,$ is a harmonic
function, say $\,u({\bf x}) = U_{\rm ext} ({\bf x})\,$, the latter
surface integral is, according to (\ref{eq:3.5}), equal to the
value of $U_{\rm ext}$ at the origin ${\rm O}$. Integrating this
result over $t$ we get
\be
\vert\, t \,\vert \ \,U_{\rm ext} ({\bf x}_{\rm O}) \ =\  \int_{\Om_t} \,
d^3 \, {\bf x} \,\ U_{\rm ext} ({\bf x}) \ , \label{eq:3.8}
\ee
which expresses the fact that the volume $\Om_t$ contained
between $\,\Si\,$ and $\,\Si_t\,$ is a homogeneous monopole, of volume
$\vert \, t\, \vert$. 

\bigskip

This construction exhausts the possible homogeneous monopoles, 
only in the simple case
where: (i) they belong to a
continuous\footnote{This basic construction method may easily be 
extended to take into
account discontinuous changes, such as, for example, the inclusion of 
additional
cavities during the growth of the monopole. There could exist,
furthermore, isolated solutions which cannot be continuously deformed.} 
family
of solutions with variable mass in which only one of the boundaries 
of the monopole is allowed to move,
and (ii) the moving boundary $\,\Si_{\hbox{\boldmath $\alpha$}}\,$ 
(where {\boldmath $\,\alpha\,$} denotes a set of parameters) coincides 
with the fixed one $\,\Si\,$ for some value of {\boldmath $\,\alpha$}, 
for which the total mass 
(i.e. volume $\,\vert\, t\, (\hbox{\boldmath $\alpha$})\,
\vert\,$) vanishes. (We assume here, as above, that the 
two boundaries $\,\Si\,$ and 
$\,\Si_{\hbox{\boldmath $\alpha$}}\,$ are connected.)

\medskip

Indeed, let us consider a ({\it a priori~} multi-parameter)
family of monopoles filling the volume $\,\Omega_{\hbox{\boldmath $\alpha$}}$
contained between the fixed boundary $\,\Si\,$ and the moving one 
$\,\Si_{\hbox{\boldmath $\alpha$}}\,$.
Let us try to perturb this monopole, at fixed volume $\,t\,$
(and therefore fixed mass), keeping the
boundary $\,\Si\,$ fixed. The perturbation of the other boundary 
$\,\Si_{\hbox{\boldmath $\alpha$}}\,$ is defined by its orthogonal displacement 
$\ \delta h =\left( 
{\bf X} (\hbox{\boldmath $\alpha$} + \delta \hbox{\boldmath $\alpha$},
\hbox{\boldmath $\xi$}) 
- {\bf X} (\hbox{\boldmath $\alpha$} , \hbox{\boldmath
$\xi$}) \right) \,.\, {\bf n}_{\Si_{\hbox{\boldmath $\alpha$}}} \, $. 
This $\,\delta h\,$ defines a single layer surface density distributed 
on $\,\Si_{\hbox{\boldmath $\alpha$}}\,$. 
Since it generates a vanishing potential outside this (connected) surface, 
it must vanish identically.
Therefore, the
only possible variations of $\,\Si_{\hbox{\boldmath $\alpha$}}\,$ 
are those in which the volume (i.e. the mass)
changes. For these, the
uniqueness of the solution of Eq.~(\ref{eq:3.5}), considered as an equation for
the harmonic measure $\,\s$, implies that $\,\delta h\,$ is necessarily equal to
$\,\s_{\Si_{\hbox{\boldmath $\alpha$}}}^{\rm O} \, \delta t 
(\hbox{\boldmath $\alpha$})$. This shows that, when one boundary $\,\Si\,$ is
fixed, the other one $\,\Si_{\hbox{\boldmath $\alpha$}}\,$ must depend only on
the single volume parameter $\,t=t(\hbox{\boldmath $\alpha$})\,$, 
and evolve according
to Eq.~(\ref{eq:3.7}). Under the further assumption that 
$\,\Si_{\hbox{\boldmath $\alpha$}}\,$ coincides with $\,\Si\,$ when 
$\,t\,(\hbox{\boldmath $\alpha$}) = 0\,$, we recover the ``thickening'' 
process of $\,\Si\,$ defined above.

\medskip

This proof of the uniqueness of the contruction of monopoles breaks down 
if we consider continuous families of solutions in which both 
the outermost boundary 
$\,\Si_{\hbox{\boldmath $\alpha$}}^{\,\rm ext}\,$, and one or 
several inner boundaries 
$\,\Si_{\hbox{\boldmath $\alpha$}}^{\,\rm int}\,$ are allowed 
to move. We shall see, 
in subsection 3.3, that the possibility to compensate arbitrary 
variations of 
$\,\Si_{\hbox{\boldmath $\alpha$}}^{\,\rm int}\,$ by appropriate 
variations of 
$\,\Si_{\hbox{\boldmath $\alpha$}}^{\,\rm ext}\,$ allows us to construct 
more general families of monopoles than the basic ones 
described by the 
thickening process of Eq.~(\ref{eq:3.7}) -- even if
this process is generalized to allow for discontinuous 
changes leading, for instance, to the inclusion of 
additional cavities during the growth process.
\medskip

Let us mention in passing that one can trivially extend our
method to constructing many sorts of {\it inhomogeneous}
monopoles. For instance, if we want to prescribe the volumic
density $\,\rho ({\bf x})\,$ within the monopole it suffices to
divide the right-hand sides of Equations (\ref{eq:3.7}) 
or (\ref{eq:3.7'})  by $\,\rho\,
({\bf x})\,$ when propagating $\,\Si_t\,$ 
($\,\vert t \vert$ now denoting the total mass of the 
monopole $\,\Om_t\,$). 
Alternatively, one can
prescribe the shape of the monopole, and define, from any
slicing of the monopole by interpolating $\,\Si_t\,$'s a
(positive) density $\,\rho \,({\bf x})\,$ so that the orthogonal
distance between $\,\Si_t\,$ and $\,\Si_{t+dt}\,$ is $\,\s_{\Si_t}^{\rm
O} / \rho\,$.

\bigskip

Instead of writing the fundamental propagation law for
homogeneous monopoles in the differential form (\ref{eq:3.7}),
one could as well write it as an ``eikonal'' type equation.
Namely, by eliminating the two surface coordinates $\,(\xi^1
,\,\xi^2)\,$ from the three embedding equations (\ref{eq:3.6}), we
can write an equation determining the position of the surface 
$\,\Si_t\,$ in the form
\be
{\bf x} \,\in \,\Si_t \ \ \Lra \ \ \vp\, ({\bf x}) = t \ . \label{eq:3.9}
\ee
In this form, the outward normal is $\,{\bf n} = \hbox{\boldmath
$\nb$} \vp / \vert \hbox{\boldmath $\nb$} \vp \vert\,$,
and the infinitesimal vectors $\,d {\bf x}\,$ connecting $\,\Si_t\,$
and $\,\Si_{t+dt}\,$ satisfy $\,dt = \hbox{\boldmath $\nb$} \vp
\cdot d {\bf x} = \vert \hbox{\boldmath $\nb$} \vp \vert \
{\bf n} \cdot d {\bf x}$. The propagation law $\ d 
{\bf x} = \s \, {\bf n} \, dt \, +$ tangent vector~ becomes
$\,\vert \hbox{\boldmath $\nb$} \vp \vert \, \s = 1\,$, i.e.
\be
(\,\hbox{\boldmath $\nb$} \vp ({\bf x})\,)^2 \ =\  (\,\s_{\Si_{\bf
x}}^{\rm O} ({\bf x})\,)^{-2} \, . \label{eq:3.10}
\ee

\medskip
Contrary to the usual eikonal 
($\,(\,\hbox{\boldmath $\nb$} \vp ({\bf x})\,)^2 = (\,n({\bf x})\,)^2\,$)
or Hamilton-Jacobi equations,
Equation (\ref{eq:3.10}) is not an ordinary local partial
differential equation for $\,\vp ({\bf x})\,$ because its
right-hand side is a non-local functional of the function $\,\vp
(\cdot )\,$ obtained by solving an elliptic problem\footnote{The
meaning of $\,\s_{\Si_{\bf x}}^{\rm O} ({\bf x})\,$ appearing on the
right-hand side of Equation (\ref{eq:3.10}) is the following:
Given the point ${\bf x}$ and the function $\vp (\cdot)$, one
determines the surface $\,\Si_{\bf x}\,$ passing through
${\bf x}$ as the set of ${\bf y} \in \Rb^N$ satisfying $\,\vp
({\bf y}) = \vp ({\bf x})\,$. Then one determines the value, at
the point ${\bf x}$, of the density of the harmonic measure on
$\,\Si_{\bf x}\,$ with respect to the fixed origin ${\rm O}$.}.
Correspondingly, the propagation equation (\ref{eq:3.7}) is
local in ``time'' but non-local in the two spatial variables $\,\xi^1
,\,\xi^2\,$. We shall briefly discuss below, in Section 5, in what functional
spaces we expect the differential equation (\ref{eq:3.7}) to admit unique
solutions (at least for $t$ varying in some interval).

\medskip

Finally, let us mention that, when working in any number of space 
dimen\-sions\footnote{It is amusing to note that, in
dimension $N=1$, any bounded object, say any (finite) reunion
of disjoint finite intervals, constitutes a homogeneous
monopole. Indeed, a general harmonic function is simply $\,u(x)
= a+bx\,$ so that the condition (\ref{eq:2.6}) holds simply by
linearity. Still, if we consider the analog of our general
construction, i.e. the growing of two intervals located
around their fixed center of gravity, one finds that the
dynamics of their growth keeps (in a simplified form) some of
the features of (\ref{eq:3.7}).} $N\geq 2$, the propagation law
(\ref{eq:3.7}) (or (\ref{eq:3.10})) takes the same form. The
only difference is that $\,\Si_t\,$ denotes a moving hypersurface
(with $\,N-1\,$ internal coordinates {\boldmath
$\,\xi\,$}). The harmonic measure $\,\s_{\Si}^{\rm O}\,$ is still
uniquely defined by (\ref{eq:3.5}), or explicitly by
$\,\s_{\Si}^{\rm O} = - \,\om_N^{\,-1} \ \part_n \, G ({\bf x} ,{\bf
x}_{\rm O})\,$ where, as mentioned in Section 2, $\,\om_N\,$ denotes
the surface of the unit sphere in $\Rb^N$ and where the Green
function has a pole $\,\sm (N-2)^{-1} \, r^{-(N-2)}\,$ when $N \ne
2$ (or $\,\ln 1/r\,$ in $\,N=2\,$) when $\,r=\vert {\bf x} - {\bf
x}_{\rm O} \vert \ra 0\,$. 

\medskip

We shall now extend the basic construction process of Equations 
(\ref{eq:3.7}) or (\ref{eq:3.10}) to more general situations.

\subsection{\sloppy Gravitationally-similar bodies and more general 
monopoles.}

We have defined, at the end of subsection 2.1, the concepts of gravitationally
equivalent and gravitationally similar bodies, which have identical
(or proportional) sequences of multipole moments. Let us show how, 
by an extension of our method, we can construct such bodies. 
A by-product of this extension will lead to more general 
monopoles than the ones obtained earlier in subsection 3.2.

\medskip

Our electrostatics considerations of subsection 3.1 admit 
the following wide generalization. Let $\,\Om\,$ denote as above a domain
with a connected boundary $\,\Si\,$. Let us consider $\,\Si\,$ 
as a grounded 
conducting surface, and introduce within it an arbitrary (fixed)
volumic charge distribution 
$\,-\,\rho_1\,({\rm Q})\,$, where ${\rm Q}$ denotes a generic point 
of $\,\Om\,$. 
This charge distribution induces on the grounded boundary $\,\Si\,$ 
a surface charge density obtained by superposing all the elementary 
charge densities
induced by point charges at the running point ${\rm Q}$.
If $\s_{\Si}^{Q}({\bf x})$ denotes the surface density (\ref {eq:3.2})
taken for a pole O located at Q (see also Eq.~(\ref {eq:71bis}) below),
the resulting surface density induced on $\,\Si\,$ by 
$\,-\,\rho_1\,({\rm Q})\,$ reads:
\be
\s_{\Si}^{\,\rho_1}\,({\bf x}) \ =
\ \int d\,V_{\rm Q}\ \, \rho_1\,({\rm Q})\ \, 
\s_{\Si}^{Q}\,({\bf x})\ \ .
\label{dist}
\ee
By construction, this surface density generates, in the exterior of $\,\Si\,$, 
the same potential as the {\it arbitrary~} volumic distribution $\,\rho_1\,$ 
contained within $\,\Si\,$. 
In other words, the surface distribution 
$\,\s_{\Si}^{\,\rho_1}\,({\bf x})\,$ is  
{\it gravitationally equivalent~} to the volumic distribution 
$\,+\ \rho_1\,({\rm Q})\,$.

\medskip

To construct an homogeneous\footnote{The construction given below 
trivially extends to inhomogeneous distributions.} 
volumic distribution $\,\rho_2\,$ equivalent to 
$\,\rho_1\,$, we can now turn on continuously the
 distribution $\,\rho_1\,$
(by considering $\,t \,\rho_1\,$, with $\,t\,$ increasing from 0 to 1), 
while growing, at the same time, successive surfaces $\,\Si_t\,$ 
from the initial surface $\,\Si\,$, in such a way that the elementary 
volume $\,d\,\Om_t\,$ of density $\,\rho_2\,$ enclosed between 
$\,\Si_t\,$ and $\,\Si_{t + dt}\,$ is 
gravitationally equivalent to the infinitesimal mass distribution 
$\ \rho_1\ dt\,$.
This new growing process is defined by the following differential 
equation (which generalizes 
Eqs.~(\ref{eq:3.7}) or (\ref{eq:3.7'})):
\be
d\,{\bf X}\ \ = \ \ \frac{1}{\rho_2} \ \,
\s_{\Si_t}^{\,\rho_1} \ \ {\bf n}_{\Si_t} \ dt\,\ ,
\label{distrib}
\ee
where $\,{\bf n}_{\Si_t}\,$ denotes the outward\footnote{To fix 
ideas we consider here an outward growth process, which has 
the advantage of being stable (as discussed later); we can define as well 
an inward growth process, by changing the sign of the right-hand side of 
Eq.~(\ref {distrib}).}
normal to $\,\Si_t\,$, and $\,\s_{\Si_t}^{\,\rho_1}\,$ the surface 
distribution
(\ref{dist}) taken for the moving surface $\,\Si_t\,$. 
The total mass of the corresponding layer $\,d\,\Om_t\,$ -- 
which by construction 
generates at the exterior of $\,\Si_t\,$ the same potential 
as the infinitesimal
distribution $\,\rho_1\ dt\,$ -- is easily verified to be
(as it should)
\begin{eqnarray}
d\,M_2&=&\ \rho_2 \ 
\int_{\,\Si_t}\, d \,\Si_t \ \ {\bf n}_{\Si_t}\,.\, d\,{\bf X}
\nonumber \\
 &=&\ 
\int_{\,\Si_t}\, d \,\Si_t\ \, \s_{\Si_t}^{\,\rho_1}\ \ \, dt \ =\ 
\int\, d\,V_{\rm Q}\ \,\rho_1\,({\rm Q}) \ \ \,dt \ =\ M_1\ dt \ \ .
\end{eqnarray}
By ending the growth process at $\,t_f=1\,$ we obtain a thickened 
version of $\,\Si\,$ which, when filled with density $\,\rho_2\,$, is 
gravitationally equivalent to the volumic distribution $\,\rho_1\,$. 
Distributions gravitationally similar to $\,\rho_1\,$ can be 
obtained by a rescaling of
$\,\rho_2$, or by ending the growth process at $\,t_f\neq 1$
(thereby generating a distribution gravitationally equivalent
to $\,t_f \rho_1\,$).

\bigskip

This construction is interesting in several respects. From a theoretical
point of view, it solves a generalization of the monopole construction 
problem by providing, for any (connected) surface $\,\Si\,$ and
any mass distribution $\,\rho_1\,$ located within $\,\Si\,$, 
an homogeneous thickened version of $\,\Si\,$ which is gravitationally 
equivalent
(or similar) to $\,\rho_1\,$. (Note that $\,\rho_1$, being totally 
arbitrary, can
correspond to a collection of disconnected bodies with arbitrary densities.)

\medskip

It also leads to the construction of more general monopoles than 
the ones obtained
by the basic growth process (\ref {eq:3.7}, \ref{eq:3.7'}), even 
when it is generalized 
by introducing discontinuities, with the inclusion of additional cavities 
during the growth. (This generalization
 can be viewed as the superposition of several 
``basic'' monopoles, all of them having the same center of mass ${\rm O}$.)

\medskip

To show this we start from an homogeneous monopole $\,{\cal M}\,$
 of center ${\rm O}$ (e.g. a solid sphere, or one of the 
monopoles constructed above). 
Let us dig arbitrarily various holes within this monopole.
 The mass distribution
$\,\rho_1\,({\rm Q})\,$ that we take off 
corresponds to a subset $\,{\cal H}\,$ of $\,{\cal M}\,$.
This density $\,\rho_1\,({\rm Q})\,$ 
taken away from the initial monopole is gravitationally equivalent, 
as we have seen, 
 to a new homogeneous body $\,\D {\cal M}\,$ 
(filled with the same density as ${\cal M}$), 
obtained by thickening,
towards the outside, the outer boundary $\,\Si\,$ of our initial 
monopole ${\cal M}$.
Altogether the resulting object 
$\ (({\cal M}- {\cal H})\ \cup \ \D  {\cal M})\ $ is a new homogeneous
 monopole,
of the same mass $\,M\,$ and center of mass ${\rm O}$ as the 
initial one. 
But it now presents a new set of holes $\,{\cal H}\,$, 
corresponding to one or several
disconnected volumes of arbitrary shapes. 

\bigskip

The construction (\ref{distrib}) is also quite interesting from a practical 
point of view. One motivation of this work is to define 
{\it differential~} accelerometers which are -- ideally -- totally 
insensitive to arbitrary external gravity gradients. 
This may be realized by using two 
gravitationally similar
bodies centered at the same point, the accelerations induced by the 
external gravity gradients, proportional to 
$\,Q_{\rho_i,\,{\bf x}_{\rm O}}^{\,i_1 i_2 \ldots i_{\ell}} \ / \ 
M_{\rho_i}\,$,
being identical.

\medskip

Given some inner body with mass distribution $\,\rho_1\,({\rm Q})\,$
(e.g. a solid straight cylinder)
and an arbitrary surface $\,\Si\,$ enclosing it, 
we can now define, from the growth process (\ref{distrib})
considered with any value of the density $\,\rho_2\,$
and any final ``time'' $t_f$, a one-parameter family
of homogeneous outer bodies 
gravitationally similar to the inner one. 
These bodies have the form  
of homogeneous thick shells built on the boundary $\,\Si\,$.
We shall see later how this construction can be extended to the case 
where $\,\Si\,$ 
has, for example, a cylindric or cylindriclike shape. 

\medskip

Before discussing in Section 5 some of the
mathematical properties of the evolution equations (\ref{eq:3.7})
or (\ref{distrib}), we shall build up our intuition
on the growing of monopoles by solving explicitly, in Section 4,
 the first step of the construction in some simple geometries.

\section{Simple examples of thin monopoles.}

\subsection{The plane and the sphere.}

The most trivial example would be the limiting case where
the domain $\,\Om\,$ is a half of $\Rb^3$, $\,\Si = \part \, \Om\,$ 
being an
infinite plane. The Green function with pole at $\,{\rm O} \in
\Om\,$ is simply $\,G({\bf x} ,{\bf x}_{\rm O}) = \vert {\bf
x} - {\bf x}_{\rm O} \vert^{-1} - \vert {\bf x} -  {\bf
x}_{\bar{\rm O}} \vert^{-1}\,$, where $\bar{\rm O}$ is the image
of ${\rm O}$ through the ``mirror'' $\,\Si\,$. Equation
(\ref{eq:3.2}) yields the surface distribution on $\,\Si\,$, 
\be
\s_{\Si}^{\rm O} \ =\  \frac{a}{2\,\pi \ r^3} \ , \label{eq:4.1}
\ee
where $\,r=\vert {\bf x} - {\bf x}_{\rm O} \vert\,$ and $\,a\,$
is the orthogonal distance between ${\rm O}$ and the plane
$\,\Si\,$. As seen from the other half-space that is the
 complement of $\,\Om\,$, the surface
distribution (\ref{eq:4.1}) generates exactly the potential
$\,\vert {\bf x} - {\bf x}_{\rm O} \vert^{-1}\,$. However, as the
asymptotic fall off of $\,\s\,$ is rather slow it does not make
sense to compute the multipole moments of the corresponding
surface distribution, nor to say that we have generated an
interesting thin monopole.

\bigskip

The second most trivial example consists of the
case where $\,\Om\,$ is a ball of radius $\,R\,$ in $\Rb^3$, $\,\Si =
\part \, \Om\,$ being a sphere. Let $\,{\rm C}\,$ be the center of the
ball and ${\rm O}$ be any point within $\,\Om\,$. The Green
function is again given by the method of images:
\be
G({\bf x} ,{\bf x}_{\rm O}) \ = \ \vert {\bf x} - {\bf x}_{\rm O}
\vert^{-1} - \lb \,\vert {\bf x} - {\bf x}_{\bar{\rm O}}
\vert^{-1} \  \quad \hbox{where} \quad \lb = R/{\rm CO} \ ,
\ee
and where $\bar{\rm O}$ is the inverse of ${\rm O}$ through
the sphere $\,\Si\,$ (i.e. $\,{\bf C}\bar{\bf O} = R^2 \ {\bf CO} / $
\break $\vert {\bf CO}\vert^2\,$). The corresponding surface
distribution is given by the well-known Poisson formula (see e.g. 
Ref.~\cite{Doob})
\be
\s_{\Si}^{\rm O} \ =\  \frac{R^2 \,-\,{\bf CO}^2}{4\pi \, R} \ \,
\frac{1}{r^3} \ , \label{eq:4.2}
\ee
which exhibits the same $\,r^{-3}\,$ dependence (where $\,r \eqv
\vert {\bf x} - {\bf x}_{\rm O} \vert\,$ with $\,{\bf x} \in \Si\,$)
as the planar distribution (\ref{eq:4.1}). (In $N$ space
dimensions the Poisson formula is simply obtained from
(\ref{eq:4.2}) by replacing $\,4\,\pi \ra \om_N\,$ and $\,r^3 \ra r^N\,$.)

\medskip

We can now consider the surface distribution
(\ref{eq:4.2}) on this sphere (of geometrical center ${\rm C}$) as 
defining 
a thin monopole (of center of mass ${\rm O}$). By
construction, it generates a Newtonian potential which is
exactly $\,\vert {\bf x} - {\bf x}_{\rm O} \vert^{-1}\,$ outside
$\,\Si\,$, its interaction energy with an external potential is
exactly that of a unit mass point located at ${\rm O}$, and all
its multipole moments (of order $\,\ell \geq 1$) around ${\rm O}$
vanish identically.

\medskip

The lesson we learn from (\ref{eq:4.1}) and (\ref{eq:4.2}) (at
least when one starts with a surface $\Si$ with a
simple shape varying only on length scales larger than the
distance between the pole ${\rm O}$ and $\,\Si\,$) is that the
growing of a monopole will consist of successive layers on $\,\Si\,$
forming bumps roughly concentrated around the points of $\,\Si\,$ 
which minimize
the distance $\,\vert {\bf x} - {\bf x}_{\rm O} \vert\,$. When
growing the monopole towards the outside, one expects the bumps to
become smeared (because their centers move away from ${\rm O}$),
and the surface $\,\Si_t\,$ to asymptotically approach a very large 
sphere of center ${\rm O}$.
On the contrary, when growing $\,\Si_t\,$ towards the inside, one
expects the bumps to start developing into fingers rapidly
advancing towards ${\rm O}$. We shall further discuss later the
instabilities associated with the inward evolution and the
smoothing character of the outward propagation.

\subsection{The thin cylinder.}

A less trivial -- and much more interesting -- example of a
 thin monopole is obtained by choosing for surface
$\,\Si = \part \, \Om\,$ a straight cylinder (of radius $R$) in $\Rb^3$.
We choose, for simplicity, the pole ${\rm O}$ on its
symmetry axis. In cylindrical coordinates $(\,\rho
,\,\vp ,\,z)$ the pole is at $\,\rho = z = 0\,$ and the equation
of $\,\Si\,$ is $\,\rho = R\,$. There are (at least) two
equivalent ways of finding the corresponding Green function.
Let us start with one before explaining the other. 

\bigskip

Writing the Green function
$\,G({\bf x}) = \frac{1}{r}\, +\, U\,(\rho ,\vp ,z)\,$, where
$\,r=\sqrt{\rho^2 + z^2}\,$, the function $\,U\,$ (manifestly
$\vp$-independent) must be harmonic within $\,\Om\,$,
\be
\frac{\part^2 \, U}{\part \, \rho^2}\, +\, \frac{1}{\rho} \,
\frac{\part \, U}{\part \, \rho}\, +\, \frac{\part^2 \,
U}{\part \, z^2} \ =\  0 \ , \label{eq:4.3}
\ee
and  equal, when $\,\rho = R\,$, to $\,-\,1/r = -\,1/\sqrt{R^2 +
z^2}\,$. Taking a Fourier transform with respect to $z$, $\,U\,(\rho
,z) = \int_{-\ify}^{+\ify}\, dk \  U_k \, (\rho) \ e^{ikz}\,$, we
find from (\ref{eq:4.3}) that $\,U_k (\rho)\,$ is proportional to a
{\it modified} Bessel function of order zero: $\,U_k (\rho) = c_k
\ I_0 (k\rho)\,$. The coefficient $\,c_k\,$ is 
obtained by writing that $\,U_k (R)\,$ is the inverse Fourier
transform of $\,U(R,z) = -\,1 / \sqrt{R^2 + z^2}\,$. The latter
transform is given by a modified Bessel function
of the second kind: $\,\int_0^{\ify} dz \, \cos kz /
\sqrt{z^2 + R^2} = K_0 (\vert k \vert R)\,$. Finally, we get
\be
G(\rho ,z) = \frac{1}{\sqrt{\rho^2 + z^2}} + U (\rho ,z) =  
\frac{1}{\sqrt{\rho^2 +z^2}} \,- \frac{1}{\pi}
\int_{-\ify}^{+\ify} dk \, \frac{K_0 (\vert k \vert
R)}{I_0 (kR)} \, I_0 (k\rho) \, e^{ikz} \ .
\label{eq:4.4}
\ee

\medskip

The corresponding surface density (\ref{eq:3.2}) can be written
(when writing also the $(\rho^2 + z^2)^{-1/2}$ term as a Fourier
integral with respect to $z$, and using the Wronskian $\,\k \,[\,
I'_0 (\k) \, K_0 (\k) - I_0 (\k) \, K'_0 (\k)\,] = 1\, $) as
\be
\s (z) \ =\  \frac{1}{2\pi \, R^2} \  \,F \left( \frac{z}{R} \right)
\  , \label{eq:4.5}
\ee
where (using the dimensionless variables $\,\z = z/R\,$ and $\,\k
=kR\,$)
\be
F (\z) \ = \ \int_{-\ify}^{+\ify}\ \frac{d\k}{2\pi} \ \frac{e^{i\k
\z}}{I_0 (\k)} \ . \label{eq:4.6}
\ee

\medskip
\medskip

A different method of getting $\,G\,$ and $\,\s\,$ is to take
advantage of the vanishing of $\,G\,(\,\rho ,\,z)\,$ at $\,\rho =R\,$ (for
any $\,z$) to expand the Green function $\,G\,(\,\rho ,\,z)\,$ 
as a Fourier-Bessel series
\be
G\,(\rho ,z)\ \,=\ \,\frac{2}{R}\ \,\sum_{\ell =1}^{\ify} \ \, 
a_{\ell}\, (z) \ 
J_0 \,(k_{\ell} \, \rho) \ , \label{eq:4.7}
\ee
where $\,k_{\ell} \, R \eqv j_{\ell} \sm 2.405,\, 5.520,\, 8.654,
\ldots\,$ 
are the successive zeroes of the Bessel function $\,J_0\,$.
Inserting this expansion into Poisson's equation
$\,\D\, G =\,-\,4\,\pi\, \d\,$ (i.e. Eq. (\ref{eq:4.3}) completed with
the source term $\,\,-\,4\,\pi \ \d (z) \, \d_{\bot} (\hbox{\boldmath
$\rho$})\,$), one gets: (i) the information that $\,a_{\ell} (z) =
A_{\ell} \, e^{-k_{\ell} \vert z \vert}\,$, and (ii), from the
strength of the delta function, the values of the coefficients
$\,A_{\ell}\,$. (More details on this, and on the properties 
of the resulting distribution function $\,F\,$, will be given elsewhere.) 
Finally, we get
\be
G\,(\,\rho ,\,z) \ \,=\ \,\frac{2}{R} \ \,\sum_{\ell =1}^{\ify} \
\,\frac{1}{j_{\ell} \ \,  J_1 (j_{\ell})^2} \ \ e^{-k_{\ell} \vert z
\vert} \ \, J_0\, (k_{\ell} \, \rho) \ , \label{eq:4.8}
\ee
in which enters the Bessel function of order one, $\,J_1\, (x) =
-\,J'_0\, (x)\,$, evaluated at the zeroes of $\,J_0 (x)\,$.
{}From (\ref{eq:4.8}) one deduces that the surface density
(\ref{eq:3.2}) can be written as in Equation (\ref{eq:4.5}), with
\be
F(\z) \ \,= \, \ \sum_{\ell =1}^{\ify} \ \,\frac{1}{J_1 (j_{\ell})} \ \,
e^{-j_{\ell}\, \vert \z \vert} \  . \label{eq:4.9}
\ee

\medskip
The identity of the different-looking results (\ref{eq:4.6}) and
(\ref{eq:4.9}) is easily verified by folding (when, say, $\,\z >
0\,$) the $\,\k\,$ contour of integration in (\ref{eq:4.6}) (after
closing it by an infinite half-circle in the upper complex $\k$
plane) around the upper imaginary axis $\,\k =ix\,$, thereby
picking up an infinite series of contributions due to residues at
the simple poles of $\,1/I_0 (ix) = 1/J_0 (x)\,$. The
Fourier-integral representation (\ref{eq:4.6}) is valid for all
values of $\,\z\,$ and makes it easy to compute numerically the
shape of the function $\,F(\z\,)$. The series (\ref{eq:4.9}) is
absolutely convergent only when $\,\z \ne 0\,$ [the slow decrease
of $\,J_1 (j_{\ell}) \sim (-)^{\ell +1} \, \ell^{-1/2}\,$ as $\,\ell
\ra \ify$ causes the series (\ref{eq:4.9}) to diverge when $\,\z
=0$], but gives a useful representation of $\,F(\z)\,$ when $\,\vert
\z \vert \gsim 1\,$ because it captures well its asymptotic fall
off. In particular, the leading behaviour for $\,\vert \z \vert >
1\,$ is given by the first term in (\ref{eq:4.9}), coming from the
first zero of $\,J_0 (x) : j_1 \simeq 2.40483$ (and $\,J_1 (j_1) \simeq
0.51915\,$). Therefore, for $\,\vert z \vert \gsim R\,$ the surface
density (\ref{eq:4.5}) is approximately given by 
\be
\s (z) \ \,\sm \ \,\frac{1.926}{2\pi \, R^2} \ \, e^{\,-\,2.405 \, \frac{\vert
z \vert}{R}} \ . \label{eq:4.10}
\ee
The function $\,F(\z)\,$ (which integrates to unity)  is represented 
graphically in Figure 1.

\medskip

By construction the surface density (\ref{eq:4.5}) laid on an
infinite straight cylinder of radius $\,R\,$ generates a $\,1/\vert
{\bf x} - {\bf x}_{\rm O} \vert\,$ potential everywhere outside
the cylinder. This monopole is of infinite extent, but, thanks
to the exponential fall off (\ref{eq:4.10}) of the density it
makes sense to consider its multipole moments of arbitrarily high
order, all the integrals $\,\int d\Si \ \s \ {\rm STF}
(x^{i_1} \ldots x^{i_{\ell}})\,$ being convergent. It is clear
(by considering, for instance, the infinite cylinder as the
limit of a finite cylinder closed by two caps) that our
construction guarantees that all the multipole moments of the
distribution (\ref{eq:4.5}) vanish identically (for $\,\ell \geq 1$). 

\bigskip

This
can be verified by explicit computations. A very simple way to
proceed (which extends to all distributions which are bounded,
or have an exponential fall off) is to consider the spatial
Fourier transform of the density distribution: $\,\hat{\rho} \,
({\bf k}) = \int d^3 \, {\bf x} \ e^{-i{\bf k} \cdot {\bf x}}
\, \rho\, ({\bf x})\,$. This is an analytic function of $\,{\bf k}\,$
which can be analytically continued to (sufficiently small)
complex values of $\,{\bf k}\,$. We define the ``inertia moment'' of order
$\,\ell\,$ of $\,\rho\, ({\bf x})$, say $I^{i_1 \ldots i_{\ell}}$, 
as the integral on
the right-hand side of Eq. (\ref{eq:2.3}) without the STF projection. 
These
inertia moments are the Taylor coefficients of the (convergent) expansion
of $\,\hat{\rho} \,({\bf k})\,$ in powers of $\,{\bf k}\,$. 
Each inertia moment
$\,I^{i_1 \ldots i_{\ell}}\,$ is equal to its trace-free projection 
$\,Q^{i_1 \ldots
i_{\ell}}\,$ plus some ``trace terms'' containing at least one Kronecker
$\,\delta^{i_a i_b}\,$. It is easily seen that a monopole ($\,Q^{i_1 \ldots
i_{\ell}} =0\,$) can be characterized by requiring that all terms of the 
Taylor
expansion of $\,\hat{\rho}\, ({\bf k})\,$ contain a factor $\,{\bf k}^2 =
\d_{ij}\, k^i k^j\,$.  In other words, a monopole is characterized by 
requiring
that the Taylor expansion of $\,\hat{\rho}\, ({\bf k})\,$ reduce to
$\,\hat{\rho}\, ({\bf 0})\,$ when the complexified $\,{\bf k}\,$ is only
restricted to satisfy $\,{\bf k}^2 = 0\,$. This general characteristic 
property
of a monopole can also be proven by taking the Fourier transform of 
the Poisson
equation  satisfied by the difference potential $V({\bf x}) = U ({\bf x}) -
\frac{M}{r}$, where $U({\bf x})$ denotes the potential generated by the
monopole (centered at the origin) and $M$ its mass. Indeed,
 $\ \triangle_x \,(U\!-\!\frac{M}{r}) 
= -\,4\,\pi\ (\rho\,({\bf x})- M\,\delta({\bf x)}) \,$, which implies 
$\,\hat{\rho}\, ({\bf k}) = \hat{\rho} \,({\bf 0}) + 
\frac{1}{4\,\pi}\ {\bf k}^2 \ \hat V ({\bf k})\,$, where $\hat V ({\bf k})$
denotes the Fourier transform of the {\it compact-support} function $V({\bf
x})$.

\medskip

For the case at
hand, the explicit calculation of $\,\hat{\rho} ({\bf k}) = \linebreak 
\int d\,\Si \
\s ({\bf x}) \ e^{-i{\bf k} \cdot {\bf x}}\,$, with $\,{\bf k} =
(\,h \cos \phi ,\, h\sin \phi ,\,k)\,$, is very easy using (\ref{eq:4.6})
and yields $\,\hat{\rho} \, ({\bf k}) = J_0\, (h) / I_0\,
(k)\,$. The fact that it reduces to $\,\hat{\rho}\, ({\bf 0}) =1\,$ when
$\,h=ik\,$ (i.e. $\,{\bf k}^2 = h^2 + k^2 =0\,$) establishes directly
the vanishing of all the multipole moments. The same
verification can be done from inserting (\ref{eq:4.9}) in
(\ref{eq:2.3}) at the price of more Bessology.

\bigskip

The important lesson we learn from this last example, by
truncating $\, \s\,$ at a finite $\,\vert z \vert > R\,$, is the
possibility to construct {\it exponentially-accurate}
approximations to exact monopoles having the shape of {\it
finite} cylinders, opened at both ends, and thickened in the
appropriate way in the middle. From (\ref{eq:4.10}) we see that the accuracy
with which the multipole moments of such a truncated thin
cylinder vanish is of order $\,e^{-j_1 \vert z \vert /R}\,$. 
More generally, one can get such exponential decreases of $\,\s\,$
at infinity, whenever, when some longitudinal dimension 
(playing the role of $z$) tends to infinity, 
the transverse dimensions remain bounded\footnote{The asymptotic 
behavior of $\,\s\,$
is then primarily determined by studying the eigenvalues of the
Schr\"{o}dingerlike equation $\,-\,\triangle_{\rm transverse}\, \psi = E \,\psi
\,$ within the interior of $\,\Si\,$,
with Dirichlet boundary conditions. The crucial criterion 
for an exponential decrease of $\,\s\,$ is the existence of a positive 
lower bound on $E$ (typically proportional to the inverse of the square 
of the transverse dimensions).}.

\medskip

Such  bodies could be very good candidates for being the ``external
cylinder'' of a differential accelerometer of the STEP type.
They are exponentially insensitive\footnote{The suppression factor
is roughly proportional to $\,\exp (-\,j_1 \, L/2R\,)$, where $\,L\,$
is the total length of the cylinder.} to external gravity
gradients, and allow easily for an ``internal cylinder''
(together with its suspension and sensing system) to be placed
inside them. 

\medskip

Alternatively we can use the generalized construction of subsection 3.3 
to define, 
for a given internal {\it straight} cylinder of finite length  -- 
with non-vanishing multipole moments -- 
a corresponding ``external cylinder'' having exactly the same reduced 
multipole moments
$\, Q^{\,i_1 i_2 \ldots i_{\ell}} \, / \, M\,$. In the thin cylinder 
approximation,
 the appropriate surface distribution $\,\s\,$ is obtained by 
generalizing the solution
(\ref{eq:4.4}-\ref{eq:4.9}) to the case where Eq.~(\ref{eq:4.3}) 
is replaced 
by $\,\triangle\,G^{\rho}\,({\bf x}) = -\,4\,\pi\ \rho\,({\bf x})\,$, 
where $\,\rho\,({\bf x})\,$
denotes the mass distribution of the internal cylinder. The 
exponential fall-off
of $\,\s\,$ will be given by the same exponential factor as in 
Eq.~(\ref{eq:4.9})
(with, in general, a different prefactor).
We shall come back, at the end, to the possible use
of this construction for Equivalence Principle experiments.

\section{Mathematical aspects of the construction of
aspherical monopoles.}

\subsection{The mathematical setting.}

We have shown in Section 3 that the problem of constructing
aspherical monopoles with a given center of
mass\footnote{In this section, the origin ${\rm O}$ used in
the construction of Section 3 is fixed from the beginning. 
We henceforth drop the
superscript ${\rm O}$ on $\,\s_{\Si_t}^{\rm O}\,$.} amounts to
solving a differential equation in the functional space
describing closed\footnote{By ``closed'', we mean more
precisely that $\,\Si\,$ is the connected boundary of a connected,
open subset of $\Rb^N$.} surfaces $\,\Si\,$. It
seems intuitively obvious that, if one starts from a
sufficiently smooth surface $\,\Si_0\,$ at ``time'' $\,t=0\,$, the
differential propagation law (\ref{eq:3.7'}), $\,d\,{\bf X} =
\s_{\Si_t} \ {\bf n}_{\Si_t} \, dt\,$, will generate, at least
for a finite interval of values of the parameter $\,t\,$ around
zero, a continuously evolving family of neighbouring surfaces
$\,\Si_t\,$. The problem of
characterizing the class of good initial data $\,\Si_0\,$ which
admit such an evolution is, actually,  highly non trivial. From the
considerations we shall now present, it is clear that (if we
demand evolution for both signs of $\,t\,$ around zero) we need to
work with {\it analytic} $\,({\rm C}^{\om})\,$ surfaces embedded in
$\Rb^N$. We have, however, not precisely determined what class
of norms one must use to guarantee that $\,\Si_0\,$ is
``sufficiently analytic''. We shall see also that the problem
of {\it outward} propagation $(t>0)$ is very different from the
{\it inward} propagation one, and allows one to start with much
less regular (say, only continuous) initial \hbox{data}\footnote{The fact 
that the evolution is stable outwards and unstable inwards
has been known for some time in a very similar two-dimensional problem
(injection of fluid in a Hele-Shaw cell;
 see e.g. Ref.\cite{elliott1}).
}.

\medskip

A first clear hint that one needs to work with analytic data
is the fact that, in two dimensions, the propagation law
(\ref{eq:3.7}) is reminiscent of the Cauchy-Riemann equations
for the mapping $\ (t,\xi) \ra {\bf X} (t,\xi) = (X,Y)\,$, 
\linebreak $\,\part X / \part t = \part Y / \part \xi\,$, 
$\ \,\part Y / \part t =-\,\part X /
\part \xi\,$, ~which can be written as \linebreak 
$\,\part \, {\bf X} (t,\xi) /
\part t =$ $ \vert \,\part \, {\bf X} / \part \xi \vert \ {\bf n}
({\bf X})\,$. Clearly the local functional $\,\vert\, \part\, {\bf X} /
\part \xi \,\vert\,$ is a much simplified version of the non-local
functional $\,\s_{\Si} ({\bf X})\,$ appearing in (\ref{eq:3.7})
which has, as we shall see, a worse sensitivity on
the regularity of the function $\,{\bf X} (\xi)\,$ than $\,\vert\,
\part \, {\bf X} / \part \xi \,\vert\,$.

\medskip

Having recognized the need to work within a (real) analytic
setting, we can abstractly formulate our problem as a
functional equation (for $\,u\,$ belonging to some functional
space) of the form
\be
\frac{d \, u(t)}{dt} \ =\ \Fc \, [\,u(t)\,] \ , \label{eq:5.1}
\ee
which is the ultimate generalization of the Cauchy-Kovaleskaya
problem. Nirenberg \cite{N72} and Nishida \cite{N77} have
proven a general theorem about the existence and uniqueness of
solutions to the general abstract non-linear
Cauchy-Kovalevskaya problem (\ref{eq:5.1}). The crucial point of
their work was to formulate the abstract analog of the
condition (in the usual Cauchy-Kovalevskaya problem) that the
functional $\,\Fc\, [\,u\,]\,$ on the right-hand side of Equation
(\ref{eq:5.1}) involve at most first-order (spatial) partial
derivatives of $u$ and be linear in them. Their formulation assumes 
that $u$
varies in a ``scale of Banach spaces'' 
$\,S=\{\, B_s ;\, 0<s \leq 1\, \}$
such that, $\,B_s \sbs B_{s'}$, when $\,s' \leq s\,$, with the norms
satisfying $\,\Vert \cdot \Vert_{s'} \leq \Vert \cdot \Vert_s\,$.
Then the crucial ``quasi-linearity'' condition (expressing the analog 
of the
linearity in first-order derivatives) is that $\,\Fc\, [\,u\,]\,$ 
must satisfy a
Lipschitz-type inequality: 
\be
\Vert \,\Fc\, [\,u\,] - \Fc\, [\,v\,]\,\Vert_{s'}\ \leq \ \,
\frac{{\rm C}}{s-s'} \ \,
\Vert\, u-v\, \Vert_s \ , \  \quad \hbox{for} \ \, s' < s \ .
\label{eq:5.2} 
\ee
Under (\ref{eq:5.2}), and some other mild assumptions, the
solution of (\ref{eq:5.1}) exists and is unique (for
sufficiently small $t$). The norms $\,\Vert \ \Vert_s\,$
appropriate to our problem are the ones standard when
dealing with analytic functions (e.g. $\,\left\Vert \,\build\,
\sum_{n=0}^{\ify} a_n \, x^n \,\right\Vert_s = \,\build
\sum_{n=0}^{\ify} \vert a_n \vert \, e^{ns}\,$) or slight
adaptations of these to deal with the embedding functions
$\ \hbox{\boldmath $\,\xi$} \in \Si\ \ra \ {\bf X} \in \Rb^N\,$.

\medskip

Actually, the functional of $\,{\bf X} (\hbox{\boldmath $\xi$})\,$
appearing on the right-hand side of the propagation equation (\ref{eq:3.7})
cannot satisfy the condition (\ref{eq:5.2}) because it contains
the {\it product} of two ``dangerous'' functionals: the normal $\,{\bf
n}_{\Si}\,$ (which depends on the first derivatives $\,\part\, {\bf X} / \part
\hbox{\boldmath $\xi$}\,$) and the harmonic measure $\,\s_{\Si}\,$. In
order to apply the abstract theorem of Refs.~\cite{N72,N77},
one should transform the original evolution equation
(\ref{eq:3.7}) into an equivalent ``quasilinear'' system that we
shall introduce later, in subsection 5.6. But, first, to understand 
the subtleties
of Equation (\ref{eq:3.7}) let us discuss the linearization of
our evolution problem, by studying the small perturbations of a monopole.

\subsection{A linear propagation equation for the perturbations of a  
mono\-pole.}

In this subsection we discuss how a monopole, 
built from an ``initial'' surface $\,\Si_0\,$, reacts to a small deformation
of this surface $\,\Si_0\,$. The deformation of $\,\Si_0\,$ is parametrized, 
at first order, by a ``height function'' 
$\,h\,(\,t\!=\!0,\,\hbox{\boldmath $\xi$})\,$
defined on $\,\Si_0\,$. This deformation, which extends to the whole 
monopole, i.e. to the whole family of surfaces $\,\Si_t$, is described -- 
still at
first order --  by a function 
$\,h\,(\,t,\,\hbox{\boldmath $\xi$})\,$ which gives the local orthogonal
displacement of the intermediate surface $\,\Si_t$. The quantity
$\,h(t,\hbox{\boldmath $\xi$})\,$ obeys a {\it linear}~ 
propagation equation, in
terms of the ``time''  evolution variable $\,t\,$ 
(i.e. the algebraic volume  of
the monopole $\,\Om_t\,$ included between the two surfaces $\,\Si_0\,$ and
$\,\Si_t\,$). We shall establish below, by different methods, this evolution
equation of $\,h\,$, and discuss, in subsection 5.3,  some general features of
its solutions.  In subsections 5.4 and 5.5 we shall explicitly solve it 
in the case of the outward propagation, and discuss, as
an illustrative example, the simple case for which the initial 
(unperturbed) monopole has a spherical symmetry of center $\rm O$. 

\medskip
The function $\,h\,(\,t,\,\hbox{\boldmath $\xi$})\,$ will be shown 
to satisfy a linear
(real ``Schr\"{o}dinger-type'' or ``heat-type'') propagation equation,
\be
\frac{\part}{\part \, t} \ h(t)\  =\  -\ \Hc (t) \ h(t) \ ,
\label{sch}
\ee
in which the (self-adjoint) operator $\,\Hc (t)\,$ is the one defining the
variation of the harmonic measure $\,\s_{\Si_t}\,$ under the perturbation
$\,h(t)\,$, according to the equation
\be
\delta \, \s_{\Si_t}\  =\  -\ \Hc (t) \ h(t) \ .
\ee
This ``Hamiltonian'' $\,\Hc \,$ is explicitly given by
\be
\Hc :\ \ \ h \ \ \ra \ \ \Hc \ h \ \eqv \  
\Nc_ {\Si_t} \, (\s_{\Si_t} \, h) \, + \, k \ \s_{\Si_t} \ h \ ,
\ee
in which $\,k\,$ denotes the mean curvature of the surface $\,\Si_t\,$ 
(i.e. $\,\frac{1}{R_1}+\frac{1}{R_2}\,$, in three dimensions), 
and $\ \Nc_{\Si_t}\ $
is the ``Neumann operator'' which associates, to every continuous 
function $\,(u)\,$
on $\,\Si_t \,$, the value $\,(\,\part_n \,\tilde u\,)\,$ 
of the normal derivative 
of its harmonic extension (within the inside of $\,\Si_t$).
 (This propagation equation of the deformation $\,h\,$, incidentally, 
must admit as a solution $\,\s_{\Si_t}\,$ itself, since 
$\,h(t)= \s_{\Si_t}\, \delta t\,$  corresponds to an obvious 
and rather trivial  deformation of the monopole associated with an 
infinitesimal
shift $\delta t$ in the variable $\,t\,$.)
The essential positivity of this Hamiltonian (meaning that it can be made
positive by adding a constant) will imply that the {\it outward} ($t>0$)
propagation of the  perturbation $\,h\,$ will be well-defined and smoothing for
all $\,t\,$,  as we shall discuss in subsections 5.3 and 5.4. We shall also 
solve explicitly
this propagation equation (\ref{sch}), in the case of the 
{\it outward} propagation.

\bigskip

Let us now assume, after this general presentation, 
that we know some exact solution of the evolution
problem (\ref{eq:3.7}), defining a monopole. 
(At this stage, the only exact solution
we can write down easily corresponds to the trivial case where 
$\,\Si_0\,$ is a
sphere {\it and}~ the origin ${\rm O}$ is chosen at its center,
 the $\,\Si_t\,$'s constituting concentric
spheres parametrized by the algebraic volume
between $\,\Si_0\,$ and $\,\Si_t\,$.) We consider a small
perturbation of this solution. If $\,{\bf X} (t,\hbox{\boldmath
$\xi$})\,$ denotes the embedding associated with the unperturbed
solution $\,\Si_t\,$, let $\,{\bf X}' (t,\hbox{\boldmath $\xi$}) = 
{\bf X} (t,\hbox{\boldmath $\xi$}) \,+\, \ve \, {\bf X}_1
(t,\hbox{\boldmath $\xi$})\,$, with $\,\ve\,$ some infinitesimal
parameter\footnote{Henceforth, we work to first order in
$\,\ve$, disregarding all terms of order
$\,\ve^2\,$, ... .}, denote the embedding associated with some
neighbouring solution $\,\Si'_t\,$. Note that we consider here 
variations of
end surfaces, $(\Si_0 ,\,\Si_t) \rightarrow (\Si'_0 ,\,\Si'_t)$, 
keeping fixed the
volume $\,t\,$ spanned between the surfaces.

\medskip

Modulo a ($t$-dependent) reparametrization of $\,\Si'_t\,$, the
perturbation of $\,\Si_t\,$ into $\,\Si'_t\,$ can be described by
a simple scalar quantity: the orthogonal distance between
$\,\Si_t\,$ and $\,\Si'_t\,$ (i.e. the local ``height'' of 
$\,\Si'_t\,$ with
respect to $\,\Si_t\,$), say (after division by $\,\ve$)
\be
h(t,\hbox{\boldmath $\xi$}) \ = \ {\bf n} (t,\hbox{\boldmath
$\xi$})\, \cdot\, {\bf X}_1 (t,\hbox{\boldmath $\xi$}) \ .
\label{eq:5.3} 
\ee
This scalar quantity also appears naturally if we perturb the
representation (\ref{eq:3.9}) instead of the embedding
representation (\ref{eq:3.6}). More precisely, if $\,\vp ({\bf x})
=t\,$ is the equation of the unperturbed family of surfaces
$\,\Si_t\,$, and $\,\vp' ({\bf x'}) \eqv \vp ({\bf x'}) + \ve \, \vp_1
({\bf x'}) =t\ $ that of the perturbed family $\,\Si'_t\,$, one finds
$\,h=-\,\vp_1 / \vert \hbox{\boldmath $\nb$} \vp \vert = -\,\s ({\bf
x}) \  \vp_1 ( {\bf x})\,$.

\medskip

An evolution equation for $\,h\,$ can be obtained by perturbing
(and projecting) the vectorial equation (\ref{eq:3.7}), or,
equivalently, by perturbing the (scalar) eikonal equation
(\ref{eq:3.10}). To describe the result we must first discuss
some mathematical objects entering the change of the harmonic
measure $\,\s_{\Si}\,$ under an infinitesimal variation of the
surface $\,\Si\,$.

\bigskip

Let us first introduce the ``Neumann operator'' $\,\Nc_{\Si}\,$
associated to $\,\Si = \part \, \Om\,$, where $\,\Om\,$ is some domain
in $\Rb^N$. The functions which are harmonic within $\,\Om\,$ (and
continuous on its closure $\,\bar{\Om} = \Om \,\cup \,\Si\,$) are
uniquely determined by their values on the boundary $\,\Si =
\part \, \Om\,$ (``Dirichlet problem''). If {\boldmath $\,\xi\,$}
denote some coordinates on $\,\Si\,$, any sufficiently regular
(say continuous) function $\,u\,(\hbox{\boldmath $\xi$})\,$ on $\,\Si\,$
defines a unique harmonic function $\,\tilde u\, ({\bf x})\,$ within
$\,\Om\,$ ($\D \tilde u =0$), called the ``harmonic extension'' of
$\,u(\hbox{\boldmath $\xi$})\,$. Under some mild regularity
conditions for $\,u\,(\hbox{\boldmath $\xi$})\,$ and $\,\Si\,$, the
normal derivative of $\,\tilde u\, ({\bf x})\,$ on $\,\Si\,$ exists.
Evidently, $\,[\,\part_n \, \tilde u\, ]\, (\hbox{\boldmath $\xi$})\,$
is a linear functional of $\,u\,(\hbox{\boldmath $\xi$})\,$. This
linear functional is called the Neumann operator:
\be
[\,\part_n \, \tilde u\, ]_{\Si}\  =\  \Nc_{\Si} \ u \ .
\label{eq:5.4} 
\ee

This operator $\,\Nc_{\Si}\,$ acts within the space of
(sufficiently regular) functions on $\,\Si\,$. It is a positive, elliptic 
pseudodifferential operator of order 1, which means essentially
that its kernel $\ \Nc (\hbox{\boldmath $\xi$} , \hbox{\boldmath
$\xi$}')\,$ is singular as $\,\hbox{\boldmath $\eta$} = 
\hbox{\boldmath $\xi$} - \hbox{\boldmath $\xi$}' \ra 0\,$ in a
manner such that its Fourier transform with respect to 
$\,\hbox{\boldmath $\eta$}\,$ is of order $\,\vert {\bf k} \vert\,$ as
$\,\vert {\bf k} \vert \ra \ify$ (${\bf k}$ denoting the
momentum associated with $\,\hbox{\boldmath $\eta$}$). It is
also known that the leading short-distance singularity of
$\,\Nc\,$ is given by $\,\sqrt{-\,\D_{\Si}}\,$ where $\,\D_{\Si}\,$ is the
intrinsic Laplacian on $\,\Si\,$, as defined by the inner metric
induced from the ambient euclidean metric on $\Rb^N$ (see
e.g. Ref.~\cite{Neumann}).

\medskip

By using the second Green's identity
\be
\int_{\Si} d\, \Si \ \ (\tilde u \ \part_n \, \tilde v \,-\, \tilde v
\ \part_n \, \tilde u ) \ =\ \int_{\Om} d^N \, {\bf x} \ \,(\tilde u
\ \D \, \tilde v\, -\, \tilde v \ \D \, \tilde u )\ =\ 0 \ ,
\ee
one sees that the Neumann operator is {\it self-adjoint} with
respect to the scalar product $\,(u,v) \eqv \int_{\Si} d\,\Si \
u\,v\,$ between functions on $\,\Si\,$: $\,(u,\,\Nc v) = (\Nc u,\,v)\,$.
Moreover, $\,\Nc\,$ is a {\it positive} operator, with respect to
this scalar product, because of the first Green's identity
\be
\int_{\Si} d\,\Si \ \ \tilde u \ \, \part_n \, \tilde u \ \,= \ 
\int_{\Om} d^N \,
{\bf x} \, \ \hbox{\boldmath $\nb$}\, (\tilde u \, \hbox{\boldmath
$\nb$} \, \tilde u ) \ \,= \ \int_{\Om} d^N \,  
{\bf x}\ \, [\,(\hbox{\boldmath
$\nb$} \, \tilde u )^2 + \tilde u \ \D \, \tilde u \,] \ ,
\ee
applied to harmonic functions:
\be
(\,u,\ \Nc u\,) \ \,=\  \int_{\Om} d^N \, {\bf x} \ \,
 (\hbox{\boldmath $\nb$} \,
\tilde u)^2 \ > \ 0 \ . \label{eq:5.5}
\ee
Note that the positivity of $\,\Nc\,$ has resulted from using the
{\it outward} normal in its definition (\ref{eq:5.4}).

\medskip

The operator $\,\Nc_{\Si}\,$ determines the modification of the
harmonic measure $\,\s_{\Si}^{\rm O}\,$ under an infinitesimal
variation of the surface $\,\Si\,$. Let us describe, as in
Equation (\ref{eq:5.3}), an infinitesimal variation of $\,\Si\,$ by
the orthogonal distance between $\,\Si\,$ and $\,\Si'\,$, i.e. as the
map 
\be
( \, {\bf X} (\hbox{\boldmath $\xi$})\, \in \,\Si \, )\ \ \ra \ \ ( \,
{\bf X}' ( \hbox{\boldmath $\xi$}) \ = \ {\bf X} (\hbox{\boldmath
$\xi$})\, +\, \ve \ h (\hbox{\boldmath $\xi$}) \, {\bf n} \ \,\in \,\Si'
\, ) \ . \label{eq:5.6}
\ee
 We can compute the
change $\,\d \, \s_{\Si}^{\rm O} (\hbox{\boldmath $\xi$}) =
\s_{\Si'}^{\rm O} (\hbox{\boldmath $\xi$}) - \s_{\Si}^{\rm O}
(\hbox{\boldmath $\xi$})\,$ (evaluated, following (\ref{eq:5.6}),
at two orthogonally corresponding points on $\,\Si\,$ and $\,\Si'\,$) by
varying the defining relation
\be
\tilde u ({\rm O}) \ = \ \int_{\Si} d\,\Si \,\ \s_{\Si}^{\rm O}
(\hbox{\boldmath $\xi$}) \, \ \tilde u\, (\,{\bf X} (
\hbox{\boldmath $\xi$})\,) \ , \label{eq:5.7}
\ee
in which $\,\tilde u ({\bf x})\,$ is an arbitrary harmonic
function. The variation of the left-hand side of (\ref{eq:5.7}) vanishes,
since one keeps both the origin ${\rm
O}$ and the function $\,\tilde u\, ({\bf x})\,$ fixed. The
variation of the right-hand side of (\ref{eq:5.7}) yields three
contributions. 

\medskip

The first one is due to the change of
the area element $\,d\,\Si\,$ (corresponding to a given coordinate
span $\,d^{N-1} \hbox{\boldmath $\xi$}\,$) under the orthogonal
deformation (\ref{eq:5.6}). This is described by the trace of
the second fundamental form of $\,\Si\,$.
We recall that, using some Gaussian coordinates, the metric of
$\Rb^N$ around $\,\Si\,$ can be written as $\,(\a ,\b ,\g =1,\ldots
,N-1)\,$
\be
ds^2 \ =\  dn^2 \ +\ g_{\a \b} (n,\xi^{\g}) \ d \xi^{\a} \,
d\xi^{\b} \ , \label{eq:5.8}
\ee
where $\,n=0\,$ is the equation of $\,\Si\,$ (so that $\,g_{\a \b}\,$ is
the first fundamental form of $\,\Si\,$) and where the curves
$\ \xi^{\a} = {\rm const.}\,$ are ambient geodesics (i.e. straight
lines in $\Rb^N$) {\it orthogonal} to $\,\Si\,$. The second
fundamental form (or extrinsic curvature tensor) of $\,\Si\,$ is
defined as one-half the normal (Lie) derivative of $\,g_{\a \b}\,$:
\be
k_{\a \b} (\xi^{\g}) \ \,\eqv \ \,\frac{1}{2} \ \left[\, \frac{\part \,
g_{\a \b} (n,\xi^{\g})}{\part \, n}\, \right]_{n=0} \, .
\label{eq:5.9} 
\ee
For a surface in $\Rb^3$, the diagonalization of the mixed
tensor\footnote{The mixed form of $k$ can also be defined as
the ``Weingarten'' map, $\,\nb_{\bf V} \, {\bf n} = k ({\bf
V})\,$, where $\,{\bf V}\,$ is a vector tangent to $\,\Si\,$, and
$\,[k({\bf V})]^{\a} = k_{\ \b}^{\a} \, V^{\b}\,$.} 
$\,k_{\ \b}^{\a} =
g^{\a \g} \, k_{\g \b}$ defines the two principal curvatures
$\,k_1 = 1/R_1\,$ and $\,k_2 = 1/R_2\,$. From $\,d\,\Si = (\det g_{\a
\b})^{1/2} \, d^{N-1} \, \hbox{\boldmath $\xi$}\,$ it is easy to
see that the variation of $\,d\,\Si\,$ under the orthogonal
deformation (\ref{eq:5.6}) depends only, at first order in
$\ve$, on the local value of $\,h(\hbox{\boldmath $\xi$})\,$ (but
not on its derivatives). It is
given by 
\be
d\,\Si' \ =\  d\,\Si \ \, [1\,+\,\ve \ h(\hbox{\boldmath $\xi$}) \ k(
\hbox{\boldmath $\xi$})] \ , \label{eq:5.10}
\ee
where
\be
k(\hbox{\boldmath $\xi$}) \ \eqv\  g^{\a \b} \ k_{\a \b} \ \eqv \
k_{\ \a}^{\a} \label{eq:5.11}
\ee
is the trace of the second fundamental form, also called the
``mean curvature''. In the most relevant case of a surface in
$\Rb^3$, $\,k=\frac{1}{R_1} + \frac{1}{R_2}\,$.

\medskip

Coming back to the variation of the right-hand side of
Equation (\ref{eq:5.7}), we find the following three
contributions 
\be
0 \ =\  \int_{\Si} d\,\Si \ \, [\,(\ve \, h \, k) \, \tilde u \ +\  \d
\, \s_{\Si}^{\rm O} \ \,\tilde u \ +\ \s_{\Si}^{\rm O}\ ( \ve \, h
\ \part_n \, \tilde u)\,] \ . \label{eq:5.12}
\ee
The third contribution (which comes from the variation of
$\,\tilde u \,({\bf X} (\hbox{\boldmath $\xi$})\,)$ caused by the
change (\ref{eq:5.6}) of its argument) can be rewritten in terms
of $\,\Nc_{\Si} \, u\,$. Finally, using the self-adjointness of the
Neumann operator, we can rewrite (\ref{eq:5.12}) as a linear
form in $\,u \,(\hbox{\boldmath $\xi$})\,$. The arbitrariness of $\,u\,(
\hbox{\boldmath $\xi$})\,$ then yields the looked for formula
for the ``normal'' variation of $\,\s_{\Si}^{\rm O}\,$
\be
\d \, \s_{\Si}^{\rm O} (\hbox{\boldmath $\xi$})\  =\ 
\s_{\Si'}^{\rm O} (\hbox{\boldmath $\xi$}) - \s_{\Si}^{\rm O} (
\hbox{\boldmath $\xi$})\  =\  - \ \ve \ [\,\Nc_{\Si} +k\,] \,
(\s_{\Si}^{\rm O} \, h) \ . 
\label{eq:5.13}
\ee
Here, $\ \Nc (\s h)=\int d\,\Si (\hbox{\boldmath $\xi$}') \ \Nc (
\hbox{\boldmath $\xi$} ,\hbox{\boldmath $\xi$}') \ \s (
\hbox{\boldmath $\xi$}') \ h(\hbox{\boldmath $\xi$}')\ $ is a
non-local functional of $\,h\,$, while $\,k \, \s \, h = k(
\hbox{\boldmath $\xi$}) \, \s (\hbox{\boldmath $\xi$}) \, h(
\hbox{\boldmath $\xi$})\,$ depends only on the local value of
$\,h\,$ at $\,\hbox{\boldmath $\xi$}\,$. (The operator $\,\Nc_{\Si} +k\,$
is independent from the choice of the pole ${\rm O}$;
therefore, by linear superposition, the equation (\ref{eq:5.13}) 
gives also the infinitesimal variation of the weighted harmonic measure 
$\,\s_{\Si}^{\rho_1}\,$ defined by Eq.~(\ref{dist}) and entering 
our generalized construction (\ref{distrib}).)

\bigskip

We are now (nearly) in measure of deriving an evolution
equation for the perturbed ``height'' $h(t,\hbox{\boldmath
$\xi$})$, defined by Equation (\ref{eq:5.3}). A direct
perturbation of the propagation law (\ref{eq:3.7}) yields
\be
\ve \ \ \frac{\part \, {\bf X}_1 (t,\hbox{\boldmath
$\xi$})}{\part \, t} \ \,= \,\ \s_{\Si'_t}^{\rm O} ({\bf X}') \ {\bf
n}' ({\bf X}') \,-\, \s_{\Si_t}^{\rm O} ({\bf X}) \ {\bf n} ({\bf
X}) \ .
\ee
Projecting this vectorial equation along the normal $\,{\bf n}\,$ 
yields (to
first order in $\ve$) the following expression for the
evolution of the height $\,h={\bf n} \cdot {\bf X}_1\,$
\be
\frac{\part}{\part \, t} \ h(t,\hbox{\boldmath $\xi$}) \ =\ 
\frac{\part}{\part \, t} \ ({\bf n} \cdot {\bf X}_1)\  =\ 
\s_{\Si'_t}^{\rm O} ({\bf X} + \ve \, {\bf X}_1) -
\s_{\Si_t}^{\rm O} ({\bf X}) + \frac{\part \, {\bf n}}{\part
\, t} \cdot {\bf X}_1 \, . \label{eq:5.14}
\ee
Because the displacement $\,{\bf X}_1\,$ has both a normal and a
tangential component, $\,{\bf X}_1 \eqv {\bf n} \, h + {\bf
X}_T\,$, the variation of $\,\s_{\Si}^{\rm O}\,$ appearing on the
right-hand side is the sum of two contributions: (1) the term
(\ref{eq:5.13}) due to the orthogonal displacement $\,h={\bf n}
\cdot {\bf X}_1\,$, and (2) the supplementary contribution $\,+ \,
{\bf X}_T \cdot \hbox{\boldmath $\nb$} \, \s_{\Si}^{\rm O}\,$ due
to the tangential displacement. However, the latter is precisely 
cancelled by the last
term on the right-hand side of Equation (\ref{eq:5.14}). Indeed,
the $t$-derivative of the normal field ${\bf n}
(t,\hbox{\boldmath $\xi$})$ is given by
\be
\frac{\part\,}{\part \, t} \ {\bf n} (t,\hbox{\boldmath $\xi$})
\ = \  -\, \hbox{\boldmath $\nb$}_T \, \s_{\Si_t}^{\rm O} \  ,
\label{eq:5.15} 
\ee
where $\hbox{\boldmath $\nb$}_T$ denotes the tangential
gradient (along $\,\Si_t\,$).

\medskip

One way to prove (\ref{eq:5.15}) is to notice that, by
definition of the $t$ and {\boldmath $\xi$} parametrizations,
the metric element of $\Rb^N$, $\,ds^2 = d \, {\bf X}^2\,$, has the
form, in $\,(t,\hbox{\boldmath $\xi$})\,$ coordinates,
\be
ds^2 \ \, = \ \,\s^2 (t,\xi^{\g}) \, dt^2 \ +\  g_{\a \b} (t,\xi^{\g}) \ \,
d\xi^{\a} \ d\xi^{\b} \ . \label{eq:5.16}
\ee
Writing the curvilinear components of the absolute
acceleration of the lines $\,\xi^{\g} = {\rm const.}\,$ is very
easy in the coordinate system (\ref{eq:5.16}) (only one
Christoffel symbol survives) and yields $\,(d^2 \, {\bf X} /
ds^2)_{\a} = (d \, {\bf n} / ds)_{\a} = -\ \part_{\a} \, \s /
\s\,$, which is equivalent to (\ref{eq:5.15}).

\medskip

Altogether, using (\ref{eq:5.13}) we get from (\ref{eq:5.14})
the simple evolution law
\be
\frac{\part\,}{\part \, t} \ h(t,\hbox{\boldmath $\xi$}) \ =\ 
-\ [\,\Nc + k\,] \, (\s \, h) \ . \label{eq:5.17}
\ee
Apart from $\,h\,$, all quantities in (\ref{eq:5.17}) refer to
the unperturbed embedding $\,{\bf x} = {\bf X} (t,\hbox{\boldmath
$\xi$})\,$ and its associated family of surfaces $\Si_t$. The
explicit expression of the right-hand side of (\ref{eq:5.17})
may be written as follows:
\be
-\int d^{N-1} \, \hbox{\boldmath $\xi$}' \
\sqrt{g(\hbox{\boldmath $\xi$}')} \ \Nc (\hbox{\boldmath
$\xi$} , \hbox{\boldmath $\xi$}') \ \s (\hbox{\boldmath
$\xi$}') \ h(t,\hbox{\boldmath $\xi$}')\ \,
-\ k(\hbox{\boldmath
$\xi$}) \ \s (\hbox{\boldmath $\xi$}) \ h(t, \hbox{\boldmath
$\xi$}) \ ,
\ee
where $\,g=\det g_{\a \b}\,$. The evolution law (\ref{eq:5.17})
applies also to the perturbation of the generalized growth process 
(\ref{distrib}) leading to the construction of gravitationally-equivalent 
bodies, if we interpret $\,\s\,$ as the weighted harmonic measure 
$\,\s_{\Si}^{\rho_1}\,$ of Eq.~(\ref{dist}).
\medskip

\medskip

An alternative way of deriving (\ref{eq:5.17}) consists of
perturbing the (scalar) eikonal equation (\ref{eq:3.10}).
Inserting 
\be
\vp' ({\bf x})\  = \ \vp ({\bf x}) \,+\, \ve \, \vp_1 ({\bf x}) \quad \
\hbox{and} \ \quad \s_{\Si'_{\bf x}}^{\rm O} ({\bf x}) \ =\  \s
({\bf x}) \,+\, \ve \, \s_1 ({\bf x})
\ee
into (\ref{eq:3.10}) yields $\ \hbox{\boldmath $\nb$} \, \vp
\cdot  \hbox{\boldmath $\nb$} \, \vp_1\, =\, -\,\s^{-3} \, \s_1\ $. The
perturbed harmonic measure is obtained from (\ref{eq:5.13}) with
the subtlety that the height $\,h\,$ is not the expected $\ -\,\vp_1 /
\vert \hbox{\boldmath $\nb$} \vp \vert = -\,\s \, \vp_1\ $, but
instead $\ h(t,\hbox{\boldmath $\xi$}') = -\,\s (\hbox{\boldmath
$\xi$}') \, [\,\vp_1 (\hbox{\boldmath $\xi$}') - \vp_1 (
\hbox{\boldmath $\xi$})\,]\ $, the last contribution coming from
the fact that, in this formulation, the surface $\,\Si_{\bf x}\,$
is constrained to pass through the point $\,{\bf x}\,$ when $\,\vp
\ra \vp'\,$ (so that the volume parameter $\,t'\! = \!\vp' ({\bf x})
\ne t\! = \!\vp ({\bf x})\,$). The resulting equation yields back
(\ref{eq:5.17}) upon using the following expression for the
$\,t$-derivative of the harmonic measure:
\be
\frac{\part}{\part \, t} \ \s (t,\hbox{\boldmath $\xi$}) \ = \ -\,
[\,\Nc + k\,] \, (\s^2) \ . \label{eq:5.*}
\ee
The latter result follows directly by applying the general
result (\ref{eq:5.13}) to the case where $\,\Si = \Si_t\,$ and $\,\Si'
= \Si_{t+\ve}\,$ are two neighbouring surfaces within the
unperturbed family of surfaces satisfying (\ref{eq:3.7}). Note
that Equation (\ref{eq:5.*}) says that $\,h=\s\,$ is a particular
solution of the general perturbation equation (\ref{eq:5.17}).
[This is indeed, as we have already mentioned, 
a ``trivial'' solution corresponding to
considering the $\,t$-shifted embedding $\,{\bf x} = {\bf X} (t+\ve
,\hbox{\boldmath $\xi$})\,$ as a ``perturbation'' of $\,{\bf x} =
{\bf X} (t,\hbox{\boldmath $\xi$})\,$.]

\medskip

\medskip

A third, instructive, way of deriving the perturbed evolution
equation (\ref{eq:5.17}) consists of directly varying the
fundamental property defining a mono\-pole, i.e.
\be
\vert\, t \,\vert \ \tilde u\, ({\rm O}) \ =\  \int_{\Om_t} d^N {\bf
x} \,\ \tilde u ({\bf x}) \ , \label{eq:5.18}
\ee
where $\,\tilde u ({\bf x})\,$ is an arbitrary harmonic function,
and $\,\Om_t\,$ the domain (of volume $\,\vert\, t\,
\vert\,$) extending between the initial surface $\,\Si_0\,$ and
$\,\Si_t\,$. Keeping fixed the origin ${\rm O}$, the volume $\,t\,$
and the harmonic function $\,\tilde u ({\bf x})\,$, the change of
(\ref{eq:5.18}) under variations of the two end surfaces $\Si_0$ and
$\Si_t$ (both described by a normal displacement as in
(\ref{eq:5.6})) is easily seen to lead to a
``conservation''\footnote{By ``conservation'' we mean
$t$-independence, as in classical mechanics when $t$ denotes
the time.} equation $\,0 = \ve \,(I_t -I_0)\,$ where
\be
I_t \ = \ \int_{\Si_t} d\,\Si_t\ \, h (t,{\bf x}) \,\ \tilde u ({\bf x})
\label{eq:5.19}
\ee
is a surface contribution associated with the variation of the ``upper cap''.
Passing as above to the (unperturbed) variables $(t,
\hbox{\boldmath $\xi$})$, one finds easily (using the tools
given above) that the ``time'' derivative of $I_t$ can be
rewritten as
\be
\frac{d \, I_t}{dt}\  =\  \int_{\Si_t} d\,\Si_t \,\ \tilde u \ \left\{ \,
\frac{\part \, h}{\part \, t}\, +\, [\,\Nc +k\,] \, (\s \, h)\,\right\}
\ \ \ (\ =\,0\ )\  , \label{eq:5.20}
\ee
thereby recovering (\ref{eq:5.17}).

\subsection{Behavior of the solutions of the propagation equation 
for the perturbation of a monopole.}

One can consider Equation (\ref{eq:5.17}) as a Schr\"odinger
equation for the ``wave function'' $h(t,\hbox{\boldmath
$\xi$})$ with respect to the imaginary time $t$ (or, directly,
as a type of ``heat equation''). To do this, it is convenient
to endow the space of functions on a surface $\Si$ with a new
scalar product differing from the canonical ``metric'' one
$ \left[\,(u,v)\eqv \int_{\Si} d\, \Si \ \,u\,v\right]\,$ used above. Let
us define the ``harmonic'' scalar product (for some given
origin ${\rm O}$) 
\be
\lgl \,u,v \,\rgl \ \,\eqv\ \, \int_{\Si} d\,\Si \,\
 \s_{\Si}^{\rm O} \,\ u\,v
\ . \label{eq:5.21}
\ee
We also define the following operators acting on the space
of functions on some surface $\Si$,
\be
\left\{   \  \begin{array}{rcl}
\Hc_0 : \ \ \ h \ \ \ra \ \Hc_0 \, h &\eqv&   \Nc (\s \, h) \, , \cr \cr
\Hc_k : \ \ \ h \ \ \ra \ \Hc_k \, h    &\eqv&  
(\Hc_0 + k \, \s) \, h = [\,\Nc +
k\,] \, (\s \, h) \ . \cr \end{array} \right. \label{eq:5.22b}
\ee
The important point is that both $\,\Hc_0\,$ and $\,\Hc_k\,$ are {\it
self-adjoint} with respect to the new scalar product
(\ref{eq:5.21}). Indeed, $\,\lgl\, u,\,\Hc_0 \, v\,\rgl = (\,\s \, u,\,\Nc
(\s \, v)\,) =$ \linebreak $ (\,\Nc (\s \, u\,),\, \s \, v) = 
\lgl \,\Hc_0 \, u,\,v
\,\rgl\,$. Another important point is that the positivity of $\,\Nc\,$
with respect to the ``metric'' scalar product, implies the {\it
positivity} of $\,\Hc_0\,$ with respect to the ``harmonic'' scalar
product (\ref{eq:5.21}). Indeed, $\lgl\, u,\,\Hc_0 \, u \,\rgl = (\,\s
\, u, \,\Nc (\s \, u)\,)$.

\bigskip

The evolution equation (\ref{eq:5.17}), 
\be
\frac{\part}{\part \, t} \ h(t)\  =\  -\ \Hc_k (t) \ h(t) \ ,
\label{eq:5.23}
\ee
appears then as the imaginary-time version of the
Schr\"odinger equation with the self-adjoint, time-dependent,
Hamiltonian $\Hc_k$. (The time-dependence of $\Hc_k$ comes
from the $\Si$-dependence of $\Nc$, $k$ and $\s$.) The formal
``Dyson'' solution of (\ref{eq:5.23}) reads
\be
h(t) \ =\  T \,\exp \, \left(-\int_0^t dt' \,\ \Hc_k (t')\right) \,
h(0) \, . \label{eq:5.24}
\ee
If the mean curvature $k$ is positive, the Hamiltonian $\Hc_k$ will be
positive. If $k$ is not positive, one can reduce the problem to a positive
Hamiltonian $\Hc'_k = \Hc_k +c$ (where $c$ is a sufficiently large 
constant) by
considering the modified variable $h' (t) = e^{-ct} \, h(t)$. 
This positivity
of the Hamiltonian makes it clear that the solution (\ref{eq:5.24}) 
will exist
for {\it positive} values of $t$ under very mild conditions on the 
regularity of
the initial data $\,h(0\,)$ as a function of the surface coordinates 
{\boldmath
$\,\xi\,$}. Like in the case of the heat equation, the evolution
(\ref{eq:5.24}) for {\it positive} $\,t\,$ will be well defined, and
{\it smoothing} for all $\,t\,$. (The explicit solution of Eq.~(\ref{eq:5.23})
given in the next subsection will show that $h(t,\hbox{\boldmath $\xi$})$ is
analytic in {\boldmath $\xi$}.)

\medskip

In other words, we expect that our evolution, when considered in the {\it
outward} direction ($t>0$) will be well defined for all $\,t\,$, 
starting from a
very general class of initial data (i.e. initial surfaces $\,\Si_0\,$) with
weak regularity properties. The surfaces $\,\Si_t\,$ generated by
the outward evolution will be very smooth (probably analytic)
even if the initial surface is barely regular. By contrast,
the situation is completely different in the case of the
evolution in the {\it inward} direction ($\,t<0\,$). In this case,
we expect that the initial data need in general to be an analytic
function of {\boldmath $\,\xi\,$}, for the solution of
(\ref{eq:5.23}) to exist at all. The domain of existence of the
solution (how ``long'' it exists for) will depend on how
analytic the initial data is. This is precisely this type of
feature that the family of norms entering the Nirenberg-Nishida
theorem \cite{N72,N77} are supposed to capture.

\subsection{Explicit solution of the propagation equation
for the perturbation of a monopole.}

We can actually write down {\it directly} (and explicitly)
the solution $\,h(t)$
of our propagation equation for the perturbation of a monopole,
 without even having to write this equation, 
in the case of the {\it outward} propagation ${(\,t>0)}$.
To do so we consider, for some given value of $t$, the variation of the
homogeneous monopole $\,\Om_t\,$ as described by the variation of both its inner
($\Si_0$) and outer ($\Si_t$)  boundaries,
expressed by the two height functions $\,h(0)\,$ and $\,h(t)\,$, respectively.

\medskip

We recall that, in this variation, the total volume $\,t\,$ spanned 
between the end
surfaces is kept fixed. Therefore, the total mass 
$\,M=\vert t \vert\,$ of the
homogeneous monopole (of unit density) does not vary, which implies 
that the
variation of the external gravitational potential 
$\,M/\vert {\bf x} - {\bf
x}_{\rm O}\vert\,$ must vanish. On the other hand, the variation of 
the external
potential is equal to the potential generated by the variation of the volumic
density of the monopole. The latter quantity is clearly made of two single
layers: a layer of algebraic thickness $\,+ \, \ve\, h(t,\hbox{\boldmath
$\xi$})\,$ on the outer boundary $\,\Si_t\,$ and a layer of algebraic 
thickness $\,- \,
\ve\, h(0,\hbox{\boldmath $\xi$})\,$ on the internal boundary $\,\Si_0$. 
The
problem is therefore, given the internal layer $\,- \, \ve\,
h(0,\hbox{\boldmath $\xi$})\,$, to determine the outer layer $\,+ \, \ve\,
h(t,\hbox{\boldmath $\xi$})\,$ which is such that the external potential
generated by these two layers vanish identically. This problem is solved by 
simple electrostatics considerations.

\medskip

Let us consider $\,\Si_t\,$ as a grounded conductor, and deposit on the
surface $\,\Si_0\,$, enclosed within $\,\Si_t\,$, a charge layer of 
surface density $\,-
\, \ve \,h (0, \hbox{\boldmath $\xi$})\,$. By definition of a grounded
conductor, the surface density induced on $\,\Si_t\,$ by the presence of 
the given
layer distribution $\,- \, \ve\, h (0, \hbox{\boldmath $\xi$})\,$ on $\,\Si_0\,$
will be the unique solution $\,+ \, \ve\, h(t,\hbox{\boldmath $\xi$})\,$ of
our problem. In mathematical terms, if $\,G_{\Si_t} ({\rm P,Q})\,$ 
denotes the Dirichlet
Green function of the domain interior to $\,\Si_t$, the potential induced at 
any
point ${\rm P}$ within $\,\Si_t\,$ by the layer 
$\,- \, \ve\, h(0)\,$ on $\,\Si_0\,$ reads
\be
v\,({\rm P})\ = \ \int_{{\rm Q} \in \,\Si_0} \ G_{\Si_t}\,({\rm P,\,Q}) 
\ \left(\,- \,\ve\ h\,(0,\, {\rm Q}) \right)\ d\,\Si_0 \ \ .
\ee
To simplify the notation we write $\,h(0,{\rm Q})\,$ 
instead of a more explicit $\,h(0,\,
\hbox{\boldmath $\xi$}_{\rm Q})\,$.

\medskip

By definition of the Dirichlet Green function this potential vanishes when 
${\rm P}
\in \Si_t\,$. The corresponding density induced on $\,\Si_t\,$ is given by 
the normal
derivative $\,\frac{1}{4\,\pi}\, \part_n \, v\,$ on $\,\Si_t\,$, so that, 
after
division by $\,\ve$,
\be
h\,(t,\, {\rm P})\ = \ \int_{{\rm Q}\, \in \,\Si_0} \ d\,\Si_0 \ \
\s_{\Si_t}^{\rm Q} ({\rm P})\ \ h\,(0,\,{\rm Q})
 \ \ . \label{solprop}
\ee
The quantity $\,\s_{\Si_t}^{\rm Q} ({\rm P})\,$ entering Eq.~(\ref{solprop}) 
is nothing but
the previously defined surface density (\ref{eq:3.2}) (``harmonic measure'')
considered for $\,\Si = \Si_t\,$ and, as in subsection 3.3,
 with pole taken at an arbitrary point ${\rm Q}$
within the interior of $\,\Si_t\,$ (instead of the fixed origin O):
\be
\s_{\Si_t}^{\rm Q} ({\rm P})\ = \ - \,\frac{1}{4\,\pi} \ \,\part_n \ 
G_{\Si_t} ({\rm P},\,{\rm Q}) \ ,
\label{eq:71bis} 
\ee
where $\,\part_n\,$ denotes the outgoing normal derivative at 
$\,{\rm P}\in \Si_t$.

\bigskip

Eq.~(\ref{solprop}) solves explicitly the problem of determining $\,h(t)\,$ in
terms of $\,h(0)$. The kernel $\,K(t,{\rm P};0,{\rm Q})\,$ 
defining the
propagation of $\,h(t)$, i.e. \linebreak$\,h(t,{\rm P}) = 
\int K (t,{\rm P};\,0,{\rm Q})\, \ h(0,{\rm Q}) \
d\,\Si_0\,$, written earlier in Eq.~(\ref{eq:5.24}) 
as a formal Dyson operator
exponential, is in fact identical to $\,\s_{\Si_t}^{\rm Q} ({\rm P})\,$.
An alternative way of proving this
result is to combine the ``conservation'' equation (\ref{eq:5.19}),
which expresses 
the invariance of the gravitational energy of the monopole 
in an external potential $\,U\,({\bf x})\,$,
\be
\int_{{\rm P}\, \in \, \Si_t} \ 
d\,\Si_t \ \ h\,(t,{\rm P}) \ \ U({\rm P})\ \,=\,\ 
\int_{{\rm Q}\, \in \, \Si_0} \ 
d\,\Si_0 \ \ h\,(0,{\rm Q}) \ \ U({\rm Q})\ ,
\label{eq:72bis}
\ee
with the
property (\ref{eq:3.5}) of the harmonic measure $\,\s_{\Si}^{\rm Q} 
({\rm P})\,$ of 
determining the unique solution
of the Dirichlet problem -- i.e. the harmonic extension of data given on a
surface $\,\Si$. More precisely, Eq.~(\ref{eq:3.5}), written by changing 
$\,{\rm O}
\ra {\rm Q}\,$, allows us to express 
\be
U\,({\rm Q})\  =\ \int_{{\rm P}\, \in\,\Si_t} \ d\,\Si_t \ \ 
\s_{\Si_t}^{\rm Q} ({\rm P})\ \ U\,({\rm P}) \label{solh} \ ,
\label{harmext}
\ee
if ${\rm Q}$ is in the interior of $\,\Si_t\,$ and $\,U\,$ 
is harmonic within $\,\Si_t$.
Combining these two equations, and using the arbitrariness of the boundary data
$\,U({\rm P})\,$ on $\,\Si_t\,$, precisely 
yields the propagation equation (\ref{solprop})
of $\,h(t)\,$. 
This argument based on the 
duality (\ref{eq:72bis}) between the perturbation $\,h\,$ and 
the external potential $\,U\,$ makes it clear why the kernel solving
the (outward) propagation of $\,h(t)\,$ (Eq.~\ref{solprop}) 
is the {\it adjoint} of the kernel solving
the (inward) harmonic extension problem (Eq.~\ref{harmext}).

\medskip

If the unperturbed family of surfaces $\,\Si_t\,$ is analytic, the Green function
$\,G_{\Si_t} ({\rm P},{\rm Q})\,$ (which solves an elliptic problem) and its 
normal derivative, proportional to $\,\s_{\Si_t}^Q ({\rm P})\,$, 
will be analytic in all their arguments. As a consequence, the
outward propagated height $\,h(t,{\rm P})\,$, given by the integral 
(\ref{solprop})
(where the points $\,{\rm P}\,$ and $\,{\rm Q}\,$ stay always separate 
when $\,t\not= 0$) will be,
when $\,t>0$, analytic in its arguments under very weak regularity assumptions
for the initial height $\,h(0,{\rm Q})\,$\footnote{Note however that if 
$\,h(0)\,$ has
sharp variations  in a given region of $\,\Si_0\,$ (e.g. if it is angulous or
even discontinuous) $\,h(t)\,$ is expected to vary rapidly in the regions of
$\,\Si_t\,$  which are close to this region of $\,\Si_0\,$. Furthermore,
$\,h(t)\,$  may even present irregularities, reflecting those of $\,h(0)\,$, at
points  $\,\hbox{\boldmath $\xi$}\,$ which are common to $\,\Si_0\,$ and
$\,\Si_t\,$. This situation is  expected to occur, as discussed later in Section
6, whenever the surface $\,\Si_0\,$ presents sharp angles ($\beta < \pi / 2$)
protruding towards its exterior. These points are expected to remain, at least for
some time, fixed  points of the successive surfaces $\,\Si_t\,$ built from
$\,\Si_0\,$, both in the cases of outward and inward growths.}. The kernel
$\,\s_{\Si_t}^{\rm Q} ({\rm P})\,$ being positive, an initial disturbance which is
everywhere positive on $\,\Si_0\,$ is propagated to an everywhere 
positive $\,h(t,{\rm P})\,$
on $\,\Si_t$. Furthermore, even if the change $\,h(0,{\rm Q})\,$ has 
compact support on $\,\Si_0$, the
propagated change $\,h(t,{\rm P})\,$ will, in general, be spread all 
over $\,\Si_t$.
Finally, the property (\ref{eq:3.3}) of the kernel $\,\s\,$ ensures
that the integral of $\,h\,$ over $\,\Si\,$ is preserved by the ``time-evolution'',
which expresses that the volume of the monopole is kept constant (and also
follows from taking $\,U=$ const. in Eq.~(\ref{eq:72bis})).

\medskip

The kernel representation (\ref{solprop}) of the solution of the outward
propagation problem does not help for solving the inward propagation one. By
reversing the arguments above, it is clear that the inward propagation of a 
perturbation $\,h(0)\,$ makes
sense only if this perturbation is analytic in the spatial
variables. In cases where, given an analytic $\,h(0)$, the integral
(\ref{solprop}) can be explicitly performed to give some well-defined analytic
function $\,h(t, \hbox{\boldmath $\xi$})\,$ for $\,t>0$, its analytic 
continuation to
negative values of $\,t\,$ would, when it is possible, define the (unique)
inward-propagated perturbation.

\medskip

Most of the results of this section hold, {\it mutatis mutandis}, 
when considering the perturbation of a body which is gravitationally-equivalent 
to some {\it fixed~} distribution $\,\rho_1\,({\rm Q})\,$. The perturbation 
may be caused, for example, by the variation of some inner boundary, and 
compensated by 
the appropriate variation of the outermost boundary. 
There is no difficulty in adapting 
the above reasoning to this case, as well as to the perturbation of 
the more general
(e.g. ``Swiss-cheese-like'') monopoles constructed at the end of 
subsection 3.3.

\subsection{\sloppy The perturbation of a spherical monopole.}

An illustrative example of the propagation of perturbation is the
case of a spherically-symmetric monopole $\Omega_t$, i.e. the volume contained
between two concentric spheres of radii $\,R(0) \equiv R_0\,$ and $\,R(t)\,$. 
The
center of gravity ${\rm O}$ must be the center of spherical symmetry. The
parameter $\,t\,$ is related to the radii by 
$\,t=\frac{4\pi}{3} (R^3 (t)-R_0^3)\,$. If
of the point
$\,{\rm P}\,$ of $\,\Si_t\,$ at which we evaluate $\,h(t)$,
 and ~{\bf Q}~ those of the point $\,{\rm Q}\,$
at which the initial perturbation of $\,\Si_0\,$ is located, the kernel 
solving the
outward propagation problem ($t>0$) is the Poisson kernel (\ref{eq:4.2}) with
pole at ${\rm Q}$, i.e.
\be
\s_{\Si_t}^{\rm Q} ({\rm P}) \ =\ \frac{R^2 (t) - R_0^2}{4\pi \, R(t)} 
\ \,\frac{1}{\vert {\bf P} - {\bf Q} \vert^3} \  \ ,\label{eq:73}
\ee
which tends asymptotically, for large $\,t$'s, towards a uniform distribution 
$\,\s \apx \frac{1}{4\,\pi\ R^2(t)}\,$, smoothing out all initial disturbances 
of $\,\Si_0\,$.

\medskip

An instructive way to see the effect of the propagation (especially in the
inward direction for which it would be incorrect to use the kernel
(\ref{eq:73})) is to solve directly the ``Schr\"odinger'' equation
(\ref{eq:5.23}). For the case at hand, the harmonic measure is $\,\s_{\Si_t} =
(4\pi \, R^2)^{-1}\,$, the curvature $\,k=2/R\,$  and the 
Neumann operator $\,\Nc
= \hat{\ell} /R\,$ where $\,\hat{\ell}\,$ is the operator acting on
functions on the sphere  as does $\ell$ on the spherical
harmonics $\,Y_{\ell m} (\t ,\vp)\,$: $\,\hat{\ell} \, Y_{\ell m} =
\ell \, Y_{\ell m}\,$. 
The explicit form of the evolution equation 
(\ref{eq:5.23}) reads 
\be
\frac{d}{dR} \ h\,(\,R,\,\t ,\,\vp) \ \,=\ \, -\  
\left[\ \frac{\hat{\ell}+2}{R}
\ \, h \ \right] \, (\,R,\,\t ,\,\vp) \ \ . \label{eq:5.25}
\ee
Its solution, given the ``initial''
perturbation of $\,\Si_0\,$ described by the orthogonal displacement
$\,h(R_0 ,\t ,\vp)\,$, is as follows. If we decompose $\,h\,$ in
spherical harmonics, 
$\ h(R,\t ,\vp) = \ \build \sum_{\ell,m}^{} \, h_{\ell m} (R)
 \ Y_{\ell m} (\t ,\vp)\,$, we have
\be
h_{\ell m}^{}\,(R)\  = \ \left( \frac{R_0}{R}\right)^{\ell +2} \,
h_{\ell m} \,(R_0) \ \  . \label{eq:5.26}
\ee

Equation (\ref{eq:5.26}) shows very explicitly the difference
between the outward and the inward evolutions. In the outward
case $(R>R_0)$, the high-harmonic fluctuations of the initial
deformation are {\it exponentially} damped as \linebreak 
$\exp \, [-(\ell
+2) \, \ln (R/R_0)]\,$. This is a typical analyticizing
transformation. In the inward case $(R<R_0)$, the
high-harmonics are enlarged by an exponential factor $\,\exp \,
[+(\ell +2) \, \ln (R_0 /R)]\,$. Therefore the solution
will exist (i.e. the spherical harmonics series will converge)
only if the initial data were such that $\,h_{\ell m} (R_0)\,$ was
an exponentially decreasing function of $\,\ell\,$, which means
that the initial shape must be analytic. 
This illustrates the general properties of the
solutions $\,h(t)\,$, that we have discussed in subsections 5.3 and 5.4.

\bigskip

Evidently, one does not need the full apparatus of the Neumann
operator, etc., to get Equation (\ref{eq:5.25}) and its solution
(\ref{eq:5.26}). Let us briefly indicate how one can directly
derive these results (generalized to any number of spatial dimensions $N$).
Consider a system of polar coordinates $(r,\hbox{\boldmath
$\t$})$ [where $\hbox{\boldmath $\t$} = (\t_1 ,\ldots
,\t_{N-1})$ denote standard angular coordinates on the unit
sphere in $\Rb^N$] centered on the origin ${\rm O}$ which is
to be the center of mass (and center of gravity). A would-be
monopole having the topology of the domain contained
between two nested spheres can be described by the
polar equations of the inner $(\Si_0)$ and outer
$(\Si_1)$ boundaries, namely $\,r\!=\!\rho_0 (\hbox{\boldmath
$\t$})\,$ and $\,r\!=\!\rho_1 (\hbox{\boldmath $\t$})\,$, respectively.

\medskip

Let $\,Y_{\ell {\bf m}} (\hbox{\boldmath $\t$})\,$ denote a basis
of spherical harmonics in $N$ dimensions (with $\,N\!-\!2\,$
``magnetic'' quantum numbers $\,{\bf m}\,$; in $\,N\!=\!2\,$, $\,{\bf m}\,$
reduces to a $\,\pm 1\,$ discrete index to distinguish $\,e^{i\ell
\t}\,$ from $\,e^{-i\ell \t}\,$, or $\,\cos \ell \t\,$ from $\,\sin \ell
\t\,$). The multipole moments of the would-be monopole (with
unit density) read
\be
Q_{\ell {\bf m}} \ =\  \int_{\Si_0}^{\Si_1} d^N {\bf x} \ \
r^{\ell} \ \, Y_{\ell {\bf m}} (\hbox{\boldmath $\t$}) \ .
\label{eq:5.27}
\ee
Replacing $\,d^N \, {\bf x} = r^{N-1} \ dr \, d \Om_{\t \!\!\!
\t}\,$ where $\,d\Om_{\t \!\!\! \t}\,$ is the area element on the
unit sphere in $\Rb^N$, and using the polar equations of the boundaries 
$\,\Si_0\,$ and $\,\Si_1\,$, we get
\be
Q_{\ell {\bf m}} \ =\  \int d\Om_{\t \!\!\! \t} \ \,Y_{\ell {\bf
m}} (\hbox{\boldmath $\t$}) \ \,\frac{1}{\ell +N} \ \, [\ \rho_1
(\hbox{\boldmath $\t$})^{\ell +N} - \rho_0
(\hbox{\boldmath $\t$})^{\ell +N}\ ] \ . \label{eq:5.28}
\ee
In the case where $\,\Si_0\,$ and $\,\Si_1\,$ are perturbed spheres
of polar equations 
$\,\rho_A (\hbox{\boldmath $\t$}) = R_A + \ve \, h (R_A
,\hbox{\boldmath $\t$})\,$ (with $\,A=0,1\,$, and $\,\ve \ll 1\,$), one
finds, to first order in $\,\ve$, and for $\,\ell \geq 1$,
\be
Q_{\ell {\bf m}} \ =\  \ve \ [\ R_1^{\ell +N-1} \ h_{\ell {\bf
m}}^* (R_1)\, -\, R_0^{\ell +N-1} \ h_{\ell {\bf m}}^*
(R_0)\ ] \ , \label{eq:5.29}
\ee
where $\ h_{\ell {\bf m}} (R_A) = \int d \Om_{\t \!\!\! \t} \
Y_{\ell {\bf m}}^* (\hbox{\boldmath $\t$}) \ \,h(R_A
,\hbox{\boldmath $\t$})\ $ is the $\,\ell
{\bf m}\,$ spherical harmonics component of $\,h(R_A
,\hbox{\boldmath $\t$})\,$. Equating $\,Q_{\ell {\bf m}}\,$ (for
$\,\ell \geq 1\,$) to zero gives
\be
h_{\ell {\bf m}} \,(R_1) \  = \ \left( \frac{R_0}{R_1} \right)^{\ell
+N-1} \ h_{\ell {\bf m}}\, (R_0) \ , \label{eq:5.30}
\ee
which is the $N$-dimensional generalization of (\ref{eq:5.26}).

\subsection{Towards a rigorous mathematical treatment 
of the growth of a monopole.}

To complete this Section dealing with the mathematical aspects of
our construction, let us sketch the rewriting of the evolution
equation (\ref{eq:3.7}) describing the growth of a monopole, 
as an evolution system that we
anticipate to be amenable to a rigourous treatment along the
assumptions of the abstract theorem of Nirenberg and Nishida.
The basic idea is, like in many proofs of the
Cauchy-Kovaleskaya theorem, to transform (\ref{eq:3.7}) (whose
right-hand side is nonlinear in the ``dangerous'' first-order
terms, i.e. $\part_{\xi \!\!\! \xi} \, {\bf X}$ and the
pseudodifferential dependence hidden in $\,\s_{\Si}$) into an
equivalent ``quasilinear'' system. Essentially, such a system
is obtained by introducing new independent variables, with
corresponding evolution equations, until one obtains a closed,
``quasilinear'' system. We think that the system obtained by
adding evolution equations for $\s$, as well as for the
{\boldmath $\xi$}-derivatives of the embedding map, $\,p_{\a}^i =
\part \, X^i (t,\hbox{\boldmath $\xi$}) / \part \, \xi^{\a}\,$,
defines a ``good quasilinear'' system which should provide a
basis for proving rigourously the existence of solutions of
(\ref{eq:3.7}) in suitable functional spaces (of the type
mentioned at the beginning of this Section). This system reads
(suppressing the {\boldmath $\xi$}-dependence):

\be
\left\{  \ \  \begin{array}{ccl}
\displaystyle{\frac{\part \,
X^i}{\part \, t}} &=&\displaystyle{ \s \ n^i (p_{\a}^j)} \ \, , \cr \cr
\displaystyle{\frac{\part \, \s}{\part \, t}}& =&
\displaystyle{ - \ \left[\ \Nc_{\bf X}\, +\,
k\left( {\bf p} ,\frac{\part \, {\bf p}}{\part \, \xi^{\a}}
\right)\ \right] \, (\s^2)  \ \, ,  }  \cr \cr
\displaystyle{\frac{\part \, p_{\a}^i}{\part \, t} }& =&\displaystyle{
\frac{\part \, \s}{\part \, \xi^{\a}} \ n^i ({\bf p}) \,+\, \s \
\frac{\part \, n^i}{\part \, p_{\b}^j} \ \frac{\part \ 
p_{\b}^j}{\part \, \xi^{\a}} \, \ . }  \cr 
\end{array} \right.  \label{eq:5.31}
\ee

\medskip

The important point is that, here, $\,{\bf n}\,$ and $\,k\,$ no
longer denote the functionals of the embedding $\,(t,\xi^{\a})
\ra X^i (t,\xi^{\a})\,$ defined above, but some explicit
(rational) functions of $\,p_{\b}^j\,$ and $\,\part \, p_{\b}^j /
\part \, \xi^{\a}\,$ obtained by replacing $\,\part \, X^i /
\part \, \xi^{\a}\,$ by $\,p_{\a}^i\,$ in the original expressions
of the normal vector $\,{\bf n} \, [\part \, X]\,$ and of the mean 
curvature $\,k \, [\part \, X, \part^2 \,
X]\,$. Similarly, $\,\s\,$ no longer denotes the harmonic measure,
solution of an
elliptic problem, but only some numerical function of
{\boldmath $\,\xi\,$}. The non-locality (in {\boldmath $\,\xi\,$}) of
the system (\ref{eq:5.31}) is all contained
in the Neumann operator $\,\Nc_{\bf X}\,$ associated with the
surface ${\bf\, x\,} = {\bf X} (t,\xi^{\a})\,$. The information that
$\,\s\,$, $\,p\,$ and $\,\part \, p / \part \, \xi\,$ are to be related to
$\,\s_{\Si}^{\rm O}\,$, $\,\part \, X / \part \, \xi\,$ and $\,\part^2 \,
X / \part \, \xi^2\,$ is fed in as suitable initial conditions
taken at $\,t=0\,$. The evolution system (\ref{eq:5.31}) will 
then preserve this property.

\medskip

The reasons why we are confident that one could establish the quasi-linearity
of the system (\ref{eq:5.31}), i.e. the validity of inequalities of the type
(\ref{eq:5.2}), are the following. First, the facts that the normal $\,{\bf n}\,
({\bf p})\,$ does not depend on the derivatives of {\bf p} with respect to
{\boldmath $\xi$}, and that the mean curvature 
$\,k \,({\bf p} ,\partial\, {\bf p} /
\part \xi^{\a})\,$ is {\it linear} in the derivatives $\,\part \,{\bf p} /
\part \xi^{\a}\,$, ensure that all the algebraic terms on the right-hand
sides of the system (\ref{eq:5.31}) are quasi-linear. The only difficulty
resides in the Neumann operator term. Here, the work of Verchota
\cite{verchota} should be sufficient to prove the quasilinearity of the term
involving $\,\Nc_{\bf X}\,$. Indeed, if (using the notation of 
Ref.~\cite{verchota}) $\,K\,$ denotes the in-surface double-layer 
potential operator and
$\,K^*\,$ its adjoint, Verchota has shown that one could control the operators
$\,\frac{1}{2} + K\,$ and $\,\frac{1}{2} + K^*\,$ even when the domain 
$\,\Om_t\,$ bounded
by $\,\Si_t\,$ is Lipshitzian. Knowing that the inequalities he derived 
depend only
of the Lipshitzian constant of $\,\Om_t\,$, and that $\,\Nc\,$ can be expressed in
terms of $\,\left( \frac{1}{2} +K \right)^{-1}\,$ or $\,\left( \frac{1}{2} + K^*
\right)^{-1}\,$ and some {\it domain-independent~} 
pseudo-differential operators,
it seems clear to us that, with some work, one can establish the
quasi-linearity of the term involving $\,\Nc$. 

\medskip

The system (\ref{eq:5.31}) can also be applied to the evolution 
(\ref{distrib}) of gravitationally-equivalent bodies. The information 
that $\,\s\,$ is to be related with the weighted harmonic measure 
(\ref{dist}) is to be fed in as an initial condition taken at $\,t=0$.

\medskip

We wish to mention that the simple representation (\ref{solprop}) of
the linearized perturbation problem suggests an alternative route towards giving
a rigourous proof of the existence of exact monopoles. If one rewrites the
exact problem as the linearized problem (around the exact solution) plus
nonlinear terms, it may be possible to use the form (\ref{solprop}) to set up a
convergent iteration scheme using at each step the exact ``propagator'' of the
previous iteration. 

\medskip

Finally, let us mention that Elliott and Janovsk\'y 
have proven, for the case of two dimensions, the
existence of a unique solution to an analogous
 problem -- the injection of fluid in a Hele-Shaw cell -- 
with, however, a different boundary condition 
near the origin \cite{elliott2}.
It would be interesting to see whether their method, 
based on an elliptic variational inequality reformulation of the moving
boundary problem, can be adapted to our situation.

\subsection{The two-dimensional case.}

In the two-dimensional case one can prove the existence of particular classes
of exact solutions of our nonlinear evolution problem (\ref{eq:3.7}) by means
of conformal mapping techniques. In  $\,N\!=\!2\,$ space dimensions, 
our problem is equivalent to the
zero-surface-tension limit of the Saffman-Taylor problem \cite{bensimon}
(see also the existence theorem of Ref.\,\cite{elliott2}). As
shown, e.g. in Refs.\,\cite{pole,bensimon}, some multi-parameter
classes of exact solutions of the latter problem can be constructed by
considering suitable $t$-dependent conformal mappings of the plane. Let us
briefly discuss the application of this idea to our problem (which has
different boundary conditions than the usual Saffman-Taylor problem).

\medskip

One introduces the complex variables $\,z=x+iy\,$ 
(where $\,(x,y)\,$ are coordinates in the
two-dimensional space) and 
$\,\zeta = e^{\,-\,(G+iH)}$, where $\,G(x,y)\,$ is the Green function,
with pole at $\,{\rm O}$, 
of the simply connected
domain $\,\Om\,$ interior to some contour $\,\Si$, and 
where $\,H(x,y)\,$ is its harmonic
conjugate 
$(\,\part G / \part x = \part H / \part y ,\ \part G / \part y = -\,\part
H / \part x\,)$. $\,\z\,$ is a holomorphic function of $\,z$. 
The (Riemann) mapping
$\,z\ra \z\,$ maps in a one-to-one conformal manner the domain 
$\,\Om\,$ onto the unit
disc in the $\,\z\,$ plane, the pole ${\rm O}$ being mapped on the 
center $\,\z = 0\,$
of the disc (the logarithmic singularity of the Green function $\,G \simeq - \,
\ln \vert z - z_{\rm O} \vert \,$ in 2 dimensions ensures that $\,\z \sim 
z-z_{\rm O}\,$ near the pole $\,{\rm O}$). 
Actually, it is more convenient to consider the inverse mapping,
$\,z=f(\z)\,$, from the unit disc onto the considered domain $\,\Om$. 
The $\,t$-evolution of a family of contours $\,\Si_t = \part \,\Om_t\,$ 
can be described as
the ``dynamics'' of $\,t$-dependent mappings $\,z=f_t (\z)\,$, $\,\Om_t\,$ 
denoting the image of the unit disc under $\,f_t\,$.

\medskip

The equation 
(\ref{eq:3.7}) governing the evolution of the boundary $\,\Si_t\,$
is equivalent to enforcing the constraint
\be
{\rm Re} \  [\,\z \ \part_{\z} \, f_t \ \ \part_t \, f_t^*\,] \ =\ {\rm Im} \
[\,\part_{\t} \, f_t \ \ \part_t \, f_t^*\,] \ =\ \frac{1}{2\,\pi} 
\label{eq:80bis}
\ee
(independently of $\,\t\,$ and $\,t$)
on the boundary of the unit disc, $\,\z = e^{\,i\t}\,$. 
The three ingredients needed to establish this constraint are:

i) along the unit circle $\,\z = e^{\,i \t}$, $\,H\,$ is identical to 
$\,- \,\t\,$; 

ii) the harmonic measure $\,\s^{\rm O}_{\Si_t}\,$ is $\,-\,\frac{1}{2\,\pi}\,$ 
times the 
normal derivative of $\,G$, i.e. the tangential derivative of $\,H$; therefore 
the differential form $\,\frac {d\t}{2\,\pi}\,$ is mapped by $\,f_t\,$ into the 
harmonic measure $\,\s^{\rm O}_{\Si_t}\, dl$, where $\,dl\,$ is the line element 
along $\,\Si_t\,$;

iii) the elementary area element swept dy $\,dl\,$ when $\,t\,$ is 
increased by $\,dt\,$ is $\,d\,{\cal A} = dl\ dh = dl\ \s^{\rm O}_{\Si_t} \ dt 
= dt\ \frac {d\t}{2\,\pi} \ $ -- also equal to 
$\,\frac{i}{2} \, df_t \wedge df_t^*\,$.
This leads to Eq.~(\ref{eq:80bis}). 

\medskip

Remarkably, one can find multi-parameter classes of mappings $\,f(\z ; p_1 ,$ 
$p_2 ,\ldots ,p_n)$ such that $\,f_t (\z) \equiv f(\z ; p_1 (t) , \ldots , p_n
(t))\,$ satisfies Eq.~(\ref{eq:80bis}) when the $\,p_i (t)\,$ 
satisfy some (complex)
ordinary differential equations $\,dp_i / dt = F_i (p_1 ,\ldots ,p_n)\,$
\cite{pole,bensimon}. The $\,p_i$'s parametrize singularities (zeroes and
poles) of the derivative $\,\part f_t / \part \z\,$. In other words, one 
can reduce
the dynamics of the contours $\,\Si_t\,$ to a particular dynamics of the 
critical
points $\,p_i$'s.

\bigskip

Let us illustrate this method by a very simple example. We consider a mapping
of the form $\,z = f_t (\z) = a(t) \,\z + \frac{1}{2} \, b(t)\, \z^2\,$. 
The derivative
mapping $\,\part f_t / \part \z = a(t) + b(t)\, \z\,$ has a zero 
(and therefore a
singularity) on the unit disc when $\vert b(t) \vert = \vert a(t) \vert$. Let
us consider the simple case where $\,a(t)\,$ and $\,b(t)\,$ are real. 
The constraint (\ref{eq:80bis}) is equivalent to the two differential equations
\be \left\{ \ \ \begin{array} {c c c}
\displaystyle{ a \,\dot a \,+\, \frac{1}{2} \ b\, \dot b } &=&
\displaystyle{ \frac{1}{2\pi} }\ , \cr \cr
\displaystyle{b \,\dot a \,+\, \frac{1}{2} \ a \,\dot b } &=& 
0 \ \, . \cr \end{array} \right.
\label{eq:ii}
\ee
The second equation shows that $\,a^2\, b \equiv C\,$ is a constant. 
The solution of
the $t$-evolution is then implicitly given by
\be
\frac{1}{2} \ a^2 \,+\, \frac{1}{4} \ b^2 \ =
\ \frac{1}{2} \ a^2 \,+\, \frac{1}{4}
\ \frac{C^2}{a^4} \ =\ \frac{t}{2\pi} \,+\, {\rm const} \ . \label{eq:iii}
\ee
The outward evolution of some initial ``epicycloid'' $\,z = a_0 \, e^{i\t}\, +\,
\frac{1}{2} \, b_0 \, e^{2i\t}\,$ (with $\vert b_0 \vert < \vert a_0 \vert$)
exhibits a growth of the ``radius'' $\,a(t)\,$ and a fast decrease of the
``ellipticity'' $\,b/a\,$. The inward evolution develops a finger 
which becomes a
cusp when $\,\vert a \vert = \vert b \vert\,$, i.e. in a finite time. 
This cusp is
{\it located away from the origin} ${\rm O}$, and 
locally described by $\,y \sim (x-x_0)^{3/2}\,$. We leave to future work a more
thorough discussion of the usefulness of conformal mapping techniques in our
context.

\section{Physical considerations.}

\subsection{On the growth of aspherical monopoles.}

The examples  of Section 4 and the 
results of Section 5 give a clear picture of the
growth of a monopole\footnote{Most of the following considerations 
also apply, {\it mutatis mutandis}, to the generalized case of the 
growth (\ref{distrib}) of a ``gravitationally-equivalent'' body.}
 obtained by thickening, towards the {\it
outside}, some initial surface $\,\Si\,$. It does not matter
whether this one has a somewhat irregular shape (say a
continuous shape with edges, corners, conical points,
etc.). Generally speaking, the thickening of $\,\Si\,$
will, at first, be most important in the regions of $\,\Si\,$
located closest to the chosen origin ${\rm O}$. The evolution
stabilizes itself in that regions of fast growth recede fast
away from ${\rm O}$ and thereby slow down their evolution. The
shape of $\,\Si\,$ becomes smoother and smoother and, in the long
term, rounder and rounder. Note, however, that in cases where the initial
surface $\,\Si_0\,$ contains narrow gaps between sub-parts of $\,\Si_0\,$ (e.g.
when $\,\Si_0\,$ is obtained by bending a ``sausage-like'' surface until the
extremities get close to each other) -- or if such a situation develops 
during the evolution -- the subsequent outward evolution can develop a
singularity (``collision'' of surfaces) in a finite time.

\medskip

 The fate of edges is interesting to
consider. Consider on $\,\Si\,$ an edge with angle $\,\b\,$ (opening
towards the inside domain $\,\Om\,$ which includes ${\rm O}$). 
The harmonic measure $\,\s\,$ is, initially,
locally proportional to $\,\rho^{\,\frac{\pi}{\b}-1}\,$ \cite{Jackson}, 
where $\rho$ denotes the distance (on $\Si$) away from the edge. This
 suggests that very acute edges, with $\,\b <
\frac{\pi}{2}\,$, remain edgelike (with the same value of $\,\b\,$)
for a while, while the surface around curves out until,
eventually, the edge disappears. Less acute edges, with $\,\b > 
\frac{\pi}{2}\,$ (and even $\,\b= \frac{\pi}{2}\,$), are expected 
to disappear instantly (i.e. for
any $\,t>0$) because of the overlap (when $\,\frac{\pi}{2} < \b <
\pi$) of the growing layers or (when $\,\b > \pi$) because of
their formally infinite growth rate.

\medskip

If we consider an infinite cylinder as initial $\,\Si\,$, with
${\rm O}$ on the axis, we expect from Section 4 the
development of an axially-symmetric bulge on the outside, with
an exponentially-decreasing tail for large $\,\vert z \vert\,$, 
proportionally to
$\, e^{\,-\,\frac{j_1 \vert z \vert}{R}}\,$ 
(the exact exponent $\,j_1\,$ in
(\ref{eq:4.10}) being preserved during the evolution).
During the thickening process the
bulge should become rounder and rounder (and therefore less and less well
approximated by the zeroth-order result of Section 4, 
represented earlier in Fig.1),  
tending asymptotically toward the shape of a sphere.

\bigskip

On the other hand the growth of a monopole obtained by thickening, towards the
{\it inside}, some initial surface $\,\Si\,$ corresponds to a completely
different picture. Consider first the case where $\,\Si\,$ is
analytic (e.g. an ellipsoid or an infinite cylinder). Then the
solution of Equation (\ref{eq:3.7}) should exist and stay
analytic for some ``time'' $\,t\,$. However, there will be a
tendency for bumps, and then ``fingers'', to grow towards the
origin ${\rm O}$. These fingers will become sharper and
sharper. The process probably ends (assuming a
``sufficiently analytic'' initial shape $\,\Si\,$) by the development 
of an infinitely sharp spine at some distance of ${\rm O}$, 
or, at the latest, when the fastest growing finger reaches the origin. 
In any case, there will be a finite ``time'' (i.e. volume) $\,\vert t
\vert\,$ before the regular growth process breaks down, i.e. a maximum filling
factor for generating a monopole with a given external surface.
Even in simple cases (such as a sphere with ${\rm O}$ not being
at the center, or an ellipsoid or a cylinder) it seems very
difficult to estimate analytically this maximum filling factor.

\medskip

If the initial shape $\,\Si\,$ is non-analytic (and, say, contains
corners, edges, conical points, $\ldots$) it is not yet clear
whether the inward growth process admits any rigourous
solution. The formula given above for corners ($\,\s \pto
\rho^{\frac{\,\pi}{\b}-1}\,$) suggests that very acute corners
($\,\b \leq \frac{\pi}{2}$) will not fill, but will become more and
more deeply incrusted within the solid body. On the other
hand, a less acute corner ($\,\b > \frac{\pi}{2}$) might admit no
uniquely defined inward evolution (when $\,\frac{\pi}{2} < \b <
\pi$ one has the collision of two growing layers, while when
$\,\b > \pi$ the tip is formally expected to grow at an infinite
rate).

\medskip

In the case of small deviations from a spherical shape 
(or more generally from a surface $\,\Si$),
the results of subsections 5.4 and 5.5 show that the problem of inward
propagation is closely  akin to the problem of defining the
{\it outward} analytic extension of the inward analytic
extension of a function given on the sphere 
(or on the surface $\,\Si$). As explained in subsection 5.4,
the two problems are actually {\it dual} to each other. This
duality leads to a simple explicit relation between the two problems
in the case of deviations from a sphere. Indeed, if one
starts from a function on the unit sphere, given, say, by its
spherical harmonic expansion $\,f(\hbox{\boldmath $\t$}) = \
\build \sum_{\ell}^{} f_{\ell} (\hbox{\boldmath $\t$})\,$, it
admits a unique inward harmonic extension within the sphere,
$\,\tilde f (r,\hbox{\boldmath $\t$}) = \ \build \sum_{\ell}^{}
r^{\ell} \, f_{\ell} (\hbox{\boldmath $\t$})\,$ with $\,r<1$. This
well-defined inward harmonic extension is closely 
related\footnote{More precisely the two problems can be
identified  if we consider the function $\,\tilde f
(R, \t,\varphi) = (R/R_0)^2 \ h(R,\t,\varphi) = \Si \ (R_0 /R)^{\ell} \ 
h_{\ell m} (R_0)\ Y_{lm}(\t,\varphi)\,$, 
and replace $\,R_0 /R\,$ by $\,r\,$.} to
the result (\ref{eq:5.26}) (with $\,R_0 /R < 1\,$) giving the
well-defined outward growth of a nearly spherical initial
$\,\Si\,$. 
The problem of continuing the
extension of $\,\tilde f (r,\hbox{\boldmath $\t$})\,$ for $\,r>1\,$ is
thereby related to that of inwardly propagating a nearly spherical
$\,\Si\,$ (Equation (\ref{eq:5.26}) with $\,R_0 /R>1\,$).

\medskip
 
Because of this analogy we might expect the inward growth
problem to be mathematically ill-defined when the initial $\,\Si$
is non-analytic. Physically, the inward growth problem can make
sense if one introduces new (regularizing) physical phenomena
taking place at small distances. In fact, there is a large
literature on such inward ``Laplacian growth'' problems (with
cut off), and they lead to a complicated zoology of fractallike
structure (for an entry into the literature see
Ref.~\cite{Laplacian}). It does not seem that the resulting objects
could be of practical interest for defining useful aspherical
monopoles (or test masses).

\medskip

Let us finally remark that some interesting phenomena can take
place in presence of symmetries (as for a parallepipedic or
cubic box with the origin ${\rm O}$ at the center of
symmetry). In such a case two or more bumps or fingers may
grow at the same rate. In the case of inward growth, the
symmetry between the fingers is unstable to small
perturbations, which should lead to interesting phenomena of
spontaneous symmetry breaking.

\subsection{On the practical realization of aspherical
mono\-poles.}

The construction we have presented
could be used for actually fabricating aspherical
(homogeneous) monopoles (or gravitationally similar test masses), 
in several different ways.
\medskip

 A first type of approach\footnote{This approach applies as well to the 
construction of gravitationally-equivalent bodies. By contrast the 
electrodeposition and brownian motion methods described below 
require more work to be applied to the case of Eq.~(\ref{distrib}).} consists
in numerically computing beforehand an ideal shape, then
machining the object according to this calculated shape, using
some computer-driven machine. There are
several practical methods one can use for computing an ideal
shape. One is to follow strictly the differential
construction given in Section 3, for instance by iterating an
electrostatic code (solving the concerned Laplacian problem)
in small successive ``time'' steps. Another method (a
``spectral'' method) is to parametrize the looked for final
shape $\Si_t$ by a finite but large number of free
parameters\footnote{The ``belted cylinder'' approach
introduced by the European STEP team \cite{Speake,STEP93,L93}
 is a simple version of this approach.
The better understanding we now have for the
theoretically-optimal shapes can help to improve upon this approach
(if nothing else, by considering
suitable multi-belted cylinders with adequate tapering!).}, and
to determine them by solving a correspondingly large number of
constraints, say the vanishing of as many multipole moments as
possible. For instance, if we define the boundaries of the
monopole by polar equations (as in subsection 5.5) the multipole
moments are given by the simple formula (\ref{eq:5.28})
 (or more complicated ones if it is not enough to
introduce two polar contours $\,\rho_0 (\t ,\vp)\,$ and $\,\rho_1 (\t
,\vp)$\,). If, for instance $\,\rho_0 (\t ,\vp)\,$ is given,
 we can approximately solve for $\,\rho_1 (\t ,\vp)\,$ by
expanding it as a linear combination of $\,K\,$ basis functions 
(which may but do not
need to be spherical harmonics) and numerically solve the first
$\,K-1\,$ equations $\,Q_{\ell m} = 0\,$ ($\,\ell \geq 1$). The optimal
choice of basis functions depends on the geometry of the
problem and the expected qualitative shape of the solution.
For instance the work of Section 4 on the cylinder suggests
that one should use basis functions that capture the basic
features of the lowest-order solution, namely a height
function above (or below) the surface  $\,\rho =R\,$ written as a
function of $\,\z =\vert z \vert / R\,$ which exhibits a smooth
hump for $\,\vert \z \vert \lsim 1\,$ followed by an exponential
decay for $\,\vert \z \vert \gsim 1$.

\bigskip

A second type of approach for realizing a monopole may consist
of using electrodeposition. Under certain conditions this may
be a very direct way of realizing the inward thickening of an
initial surface $\,\Si\,$. Let us consider the space $\,\Om\,$
inside some given surface $\,\Si\,$ to be (uniformly) filled with
an electrolyte solution (of conductivity $\,\eta\,$) and let us
apply a difference of potential between the cathode $\,\Si\,$ and
an almost pointlike anode located at the origin ${\rm O}$. The
electric field $\,{\bf E}\,$ induces, in the quasi-stationary
regime, a current density $\,{\bf j} = \eta \, {\bf E}\,$,
satisfying $\,\hbox{\boldmath $\nb$}\! \cdot \!{\bf j} =0\,$
except at the origin ${\rm O}$. This current density obeys the
same differential equation as $\,{\bf E}\,$. Provided that we can
consider the surface of the growing cathode as an
equipotential, the magnitude of $\,{\bf j}\,$ on $\,\Si\,$ is
proportional to the harmonic measure $\,\s\,$. If we take for
electrolyte a metallic solution, the thickness of the metal
sheet deposited on the cathode $\,\Si\,$ per unit time, fixed by
$\,\vert {\bf j} \vert\,$, is proportional to $\,\s\,$. Assuming the
surface of the growing cathode to remain an equipotential
under conditions allowing for the stable\footnote{Many
physical instabilities, in addition to the mathematical
unstabilities mentioned above if the initial shape $\,\Si\,$ is
not smooth enough, might alas show up. In particular, in order to
avoid such instabilities we may have to ensure that the resistivity
of the electrolyte is small enough, compared to that of the
growing aggregate \cite{DKS87}, a requirement which tends to
be in conflict with the other requirement that the surface of
the growing aggregate remains an equipotential.} growth of a
dense pattern, this method of electrodeposition can provide
us with a practical mean to thicken $\,\Si\,$ precisely according
to the law (\ref{eq:3.7}).

\medskip

An alternative way of getting this
growth law 
is to have a stationary Brownian motion with a source of
particles at $\,{\rm O}\,$ and to kill it when the particles first
hit $\,\Si\,$. The statistics of the hittings on $\,\Si\,$ is precisely
given by the harmonic measure $\,\s\,$ (see e.g. Ref.~\cite{Doob}), so
that, if we conceive the particles as sticking on $\,\Si\,$ when
they hit it, we have a Monte Carlo version of
electrodeposition. This could be useful either as a
computational method, or, possibly (if one can think of a
suitable implementation), as a practical realization method.

\section{CONCLUSIONS.}

We have shown how to construct aspherical homogeneous bodies
behaving exactly as point masses, both from the point of view
of the gravitational field they generate, and from the point
of view of their interaction with external gravitational
fields. All their multipole moments (of order $\,\ell \geq 1$)
 vanish exactly. We have proven that these
gravitational monopoles cannot occupy a ``solid'' region of
space (in the sense of a solid sphere) but must possess (at
least) one internal cavity. There is a very large flexibility
in the construction of such objects (in any number of space
dimensions). The free parameters entering this construction of
monopoles in $N$ spatial dimensions (besides the value of the
uniform density) are (i) an arbitrary ``initial'' hypersurface
$\,\Si\,$ bounding a connected open subset $\,\Om\,$ of $\Rb^N$; (ii)
the arbitrary choice of the final center of mass (and ``center
of gravity'') ${\rm O}$ within $\,\Om\,$; and (within some limits)
the total volume of the constructed body.

\medskip

An extension of this method allows one to construct homogeneous bodies
which are gravitationally equivalent (in the sense of having exactly
the same infinite  
sequence of multipole moments) or gravitationally similar 
(with proportional sequences of multipoles) to any given body, 
or collection of bodies. 
There is a very large flexibility in the construction of these objects, 
given in particular the arbitrariness in the choice of the initial surface
upon which the construction process is based. By combining and 
superposing the two constructions, we can obtain swiss-cheese-like 
monopoles having 
an arbitrary number of cavities, with arbitrarily chosen shapes.

\medskip

 The basic
construction consists of thickening the arbitrary initial
hypersurface $\,\Si\,$ by depositing homogeneous layers with
infinitesimal successive heights proportional to the harmonic
measure, with respect to the fixed pole\footnote{The construction of 
gravitationally-equivalent bodies is similar, provided one replaces 
the harmonic measure $\,\s_{\Si_t}^{\rm O}\,$ by the weighted measure
$\,\s_{\Si_t}^{\rho_1}\,$ of Eq.~(\ref{dist}).}
 ${\rm O}$, of the
moving surface $\,\Si_t\,$. The natural evolution parameter for
this growing process is the ($N$-dimensional) volume contained
between $\,\Si\,$ and $\,\Si_t\,$. The topology of $\,\Si\,$ is not
restricted to be that of a sphere in $\Rb^N$; e.g. in 3
dimensions $\,\Si\,$ can be a multi-handled ``torus''. The growth
process has very different properties when performed
outwardly, or inwardly\footnote{Though
we phrased our construction in the text as being either
outward or inward, it is evident that one may want, in some
practical applications, to thicken some ``initial'' $\,\Si\,$ on
both sides at once.}, with respect to $\,\Si\,$. Our study of the 
linearized evolution
equation (a kind of imaginary-time Schr\"odinger
equation, i.e. similar to the heat equation), and of its solutions, 
shows that the
outward growth process is stable and smoothing. It is probably
well defined even if the initial hypersurface has a very
limited regularity. By contrast, the inward growth process is,
probably, mathematically well defined only for analytic
initial data. We presented the program of a rigourous
mathematical proof of the existence of aspherical monopoles without,
however, completing such a proof. We also exhibited explicitly a 
simple example of
a two-dimensional aspherical monopole.

\medskip

We indicated several practical ways of realizing gravitational
monopoles (including computer-aided machining and
electrodeposition). In particular, once one knows, from the
general construction method, that such objects exist with many
possible shapes, one can try to find numerical approximations
to some of them (as Barrett \cite{B89} did some years ago)
without necessarily following into the footsteps of the
harmonic measure construction (see Section 6.2). Though our
method generally assumes that the domain $\,\Om\,$ is bounded, it
can also generate unbounded monopoles, with cylindriclike
internal cavities, and exponentially-decreasing thicknesses at
infinity.

\medskip

Apart from its conceptual interest, our construction might be
of practical interest for optimizing the shape of test masses
in high-precision Equivalence Principle experiments, such as
STEP. By suppressing the coupling to gravity gradients, it
allows one to design, with great flexibility in the choice of
shapes, differential accelerometers, made of nested bodies,
which are insensitive to external gravitational disturbances.
Even very close disturbing masses would have no
effect\footnote{When the monopoles are concentric.} on the
differential accelerometer, which is then 
{\it ``gravitationally screened''} from
all external gravity gradients. Truncated versions of the
cylindriclike monopoles with exponentially falling off
thicknesses may define particularly useful shapes for the STEP
experiment. If their length $\,L\,$ is large enough compared to
their radius $\,R\,$ such cylinders would be
exponentially insensitive to disturbing 
masses\footnote{Our lowest-order solution -- and its successive 
iterations -- indicate
that the quality factor of the decoupling goes like 
$\,e^{\,2.405 \,L/2R}\,$,
increasing by about an order of magnitude each time the
total length $\,L\,$ is increased by $\,2R\,$.}, even if these
are very close.

\medskip

Using the extended construction method described in subsection 3.3 
we can choose
any (non-monopole) shape for some inner body, e.g a straight 
cylinder, and 
construct an enclosing (or cylinderlike) outer body which has exactly 
(or almost exactly) the  same set of
reduced multipole moments $\,Q_{\ell}\,/M\,$ as the inner body. 
This is enough to ensure that 
the differential
accelerometer they define is insensitive to all external 
gravity gradients.

\medskip

Finally our techniques may also allow
one to devise {\it multi-nested} differential accelerometers,
made of ``Russian doll'' type masses, (almost)
insensitive to external gravitational disturbances. This could be
important for conceiving and implementing more compact (and
cheaper) versions of the STEP concept. Altogether, the optimization 
of the test masses used in differential accelerometers should allow 
for a significant increase of their performances, as required 
for very precise tests of the Equivalence Principle.

\vglue .8cm

\noindent {\bf Acknowledgments:} T.D. is grateful to Stanley
Deser for a discussion which was at the origin of this work.
We thank M. Brodsky, P. Cartier, G. David, B. Derrida, E. Guyon, 
V. Hakim, S. Roux, S. Semmes and P. Touboul for informative discussions
and/or communications. We also thank a referee for bringing to our attention 
some relevant references.

\vglue 1.3cm

\vskip 1truecm

\noindent{\Large \bf Figure caption.}

\bigskip
\medskip

\noindent
{\bf Fig.~1}: Surface density for a thin cylindrical 
gravitational monopole of radius $R$, as a function of the 
longitudinal coordinate $z$. 
The function  $F\,(z/R)\,$, defined by Eqs.~(\ref{eq:4.6}) 
or (\ref{eq:4.9}),
determines the mass density \hbox{$\ \sigma\,(z) =\,\frac{(M)}{2\,\pi\, R^2} 
\ F\,(z/R)\,$}.
All  $(l \geq 1)$ multipole moments vanish identically. 
 86 \% of the mass correspond to 
$\,|\,z\,| < R\, $, ~\hbox{98.7 \%} to $\,|\,z\,| < 2\,R\,$.

\end{document}